\documentclass[journal]{IEEEtran}
\pdfoutput=1

\ifCLASSINFOpdf
\else
\fi

\usepackage[cmex10]{amsmath}
\usepackage{tikz,pgfplots}
\usepackage{graphicx}
\usepackage{placeins}
\usepackage{mfirstuc}
\usepackage{mathtools}
\usepackage{wrapfig}
\usepackage{pdfpages}
\usepackage{diagbox}
\usepackage{makecell}
\usepackage{amssymb}
\usepackage{bbm}
\usepackage[
ruled,
vlined,
boxed, 
linesnumbered, 
commentsnumbered]{algorithm2e}

\usetikzlibrary{backgrounds,shapes,arrows,positioning}
\usetikzlibrary{shapes,snakes}
\usetikzlibrary{fit,calc} %

\usepackage[colorlinks=true,bookmarks=false,citecolor=blue,urlcolor=blue]{hyperref} %
\usepackage{textcomp}
\usepackage{acronym}
\let\mathcal\undefined
\DeclareMathAlphabet{\mathcal}{OMS}{cmsy}{m}{n}

\newcommand{\textsub}[1]{\textnormal{#1}}
\newcommand{\CW}{\mathcal{C}}

\newcommand{\dmin}{d_{\textnormal{min}}}
\newcommand{\ddesign}{d_{\textup{des}}}

\newcommand{\que}{\mathord{?}}

\newcommand{\ZO}{\{0, 1\}}
\newcommand{\ZQO}{\{0, \que, 1\}}

\usepackage{bm}
\renewcommand{\vec}[1]{\bm{#1}}
\DeclareMathOperator{\DF}{\mathsf{D}}
\newcommand{\DFC}{\DF_{\textsub{C}}}
\newcommand{\BDD}{\textnormal{BDD}}
\newcommand{\EaED}{\textnormal{EaED}}
\newcommand{\dH}{\textup{d}}

\newcommand{\dnE}[1]{\operatorname{d}_{\sim{\textnormal{E}(#1)}}}
\newcommand{\Sp}{\mathcal{S}}
\newcommand{\E}{\textnormal{E}}

\newcommand{\Eb}{E_{\textnormal{b}}}
\newcommand{\No}{N_{\textnormal{0}}}
\newcommand{\deltaC}{\delta_{\textnormal{c}}}
\newcommand{\epsilonC}{\epsilon_{\textnormal{c}}}
\DeclareMathOperator{\QFunc}{Q}
\newcommand{\Topt}{T_{\textnormal{opt}}}
\newcommand{\Lc}{L_{\textnormal{c}}}
\newcommand{\Ta}{T_{\mathrm{a}}}
\newcommand{\ind}[1]{\mathbbm{1}_{\{#1\}}}

\definecolor{KITpalegreen}{RGB}{130,190,60}
\definecolor{KITcyanblue}{RGB}{80,170,230}
\definecolor{KITorange}{rgb}{.87,.60,.10}
\definecolor{tab104}{RGB}{152,78,163}
\definecolor{colorDRSD}{RGB}{255,128,0}
\definecolor{colorideal}{RGB}{224,243,248}
\definecolor{colorDRSDdeter}{RGB}{168,129,188}
\definecolor{colorDRSDplus20}{RGB}{255,102,178}
\definecolor{colorBEE-PC}{RGB}{96,96,96}
\definecolor{colorAD}{RGB}{0,102,51}

\acrodef{HDD}{hard decision  decoding}
\acrodef{SDD}{soft decision decoding}
\acrodef{TPD}{turbo product decoding}
\acrodef{BSC}{binary symmetric channel}
\acrodef{iBDD}{iterative bounded-distance decoding}
\acrodef{BDD}{bounded distance decoding}
\acrodef{SA-HDD}{soft-aided \ac{HDD}}
\acrodef{PC}{product code}
\acrodef{AD}{anchor decoding}
\acrodef{HRB}{highly reliable bit}
\acrodef{EaE}{error-and-erasure}
\acrodef{EaED}{error-and-erasure decoder}
\acrodef{SABM}{soft-aided bit marking}
\acrodef{iEaED}{iterative error-and-erasure decoding}
\acrodef{BI-AWGN}{binary input additive white Gaussian noise}
\acrodef{BCH}{Bose--Chaudhuri--Hocquenghem}
\acrodef{DRS}{dynamic reliability score}
\acrodef{DRSD}{dynamic reliability score decoder}
\acrodef{BER}{bit error rate}
\acrodef{SABM-SR}{SABM with scaled reliabilities}
\acrodef{NCG}{net coding gain}
\acrodef{GMD}{generalized minimal distance}
\acrodef{GMDD}{generalized minimal distance decoder}
\acrodef{RS}{Reed--Solomon}
\acrodef{DE}{density evolution}
\acrodef{LLR}{log-likelihood ratio}
\acrodef{BPSK}{binary phase shift keying} 
\acrodef{SNR}{signal-to-noise ratio} 
\acrodef{BEE-PC}{binary message passing based on \ac{EaE} decoding for \acp{PC}}
\acrodef{EMP}{extrinsic message passing}
\newcommand{\yimp}{\vec{y}}

\newcommand{\yone}{\vec{y}^{(\textnormal{1})}}
\newcommand{\ytwo}{\vec{y}^{(\textnormal{2})}}
\newcommand{\pone}{\vec{p}^{(\textnormal{1})}}
\newcommand{\ptwo}{\vec{p}^{(\textnormal{2})}}

\newcommand{\wone}{\vec{w}^{(\textnormal{1})}}
\newcommand{\wtwo}{\vec{w}^{(\textnormal{2})}}
\newcommand{\wi}{\vec{w}^{(i)}}
\newcommand{\yi}{\vec{y}^{(i)}}
\newcommand{\Ps}{P_{\textnormal{s}}}

\definecolor{colorDRSD3bit}{RGB}{252,174,145}
\definecolor{colorDRSD4bit}{RGB}{251,106,74}
\definecolor{colorDRSD6bit}{RGB}{165,15,21}

\definecolor{cR1}{rgb}{0,.59,.51}
\definecolor{cR2}{RGB}{162,34,35}
\definecolor{cR12}{RGB}{217,115,14}

\pgfplotsset{compat=newest} 
\begin{document}
    
    \title{Improved Soft-aided Decoding of Product Codes with Dynamic Reliability Scores}
    
    \author{Sisi~Miao,~\IEEEmembership{Student~Member,~IEEE}, Lukas~Rapp,~\IEEEmembership{Student~Member,~IEEE}, and~Laurent~Schmalen,~\IEEEmembership{Senior~Member,~IEEE}
        
        \thanks{This work was supported by the European Research Council under the European Union’s Horizon 2020 Research and Innovation Programme under Grant 101001899. Parts of this paper have been presented at the Optical Fiber Communication Conference (OFC), 2022~\cite{miao2021improved}. (Corresponding author: Sisi Miao.)}
        \thanks{The authors are with the Karlsruhe Institute of Technology (KIT), Communications Engineering Lab (CEL), 76187 Karlsruhe, Germany (e-mail: \{\texttt{sisi.miao@kit.edu, lukas.rapp3@student.kit.edu, schmalen@kit.edu\}}.}
        \thanks{Color versions of one or more figures in this article are available at https://doi.org/10.1109/JLT.2022.3201951.}
        \thanks{Digital Object Identifier 10.1109/JLT.2022.3201951}}
    
    \markboth{Accepted version, JOURNAL OF LIGHTWAVE TECHNOLOGY}%
    {MIAO \MakeLowercase{\textit{et al.}}: Improved Soft-aided Decoding of Product Codes with Dynamic Reliability Scores}
    
    \maketitle
    
    \begin{abstract}
        Products codes (PCs) are conventionally decoded with efficient iterative bounded-distance decoding (iBDD) based on hard-decision channel outputs which entails a performance loss compared to a soft-decision decoder. Recently, several hybrid algorithms have been proposed aimed to improve the performance of iBDD decoders via the aid of a certain amount of soft information while keeping the decoding complexity similarly low as in iBDD. We propose a novel hybrid low-complexity decoder for PCs based on error-and-erasure (EaE) decoding and dynamic reliability scores (DRSs). This decoder is based on a novel EaE component code decoder, which is able to decode beyond the designed distance of the component code but suffers from an increased miscorrection probability. The DRSs, reflecting the reliability of a codeword bit, are used to detect and avoid miscorrections. Simulation results show that this policy can reduce the miscorrection rate significantly and improves the decoding performance. The decoder requires only ternary message passing and a slight increase of computational complexity compared to iBDD, which makes it suitable for high-speed communication systems. Coding gains of up to 1.2\,dB compared to the conventional iBDD decoder are observed.
    \end{abstract}
    
    \begin{IEEEkeywords}
        soft-aided hard decision decoding, product codes, optical communication
    \end{IEEEkeywords}
    
    \section{Introduction}
\Acp{PC}~\cite{Elias1955} are powerful code constructions with high \acp{NCG} that can be obtained with low-complexity decoders suitable for e.g., high-speed optical fiber communications. A PC codeword is a 2-D array where every row and column is protected by a component code, which is typically a \ac{RS} code or a \ac{BCH} code.
In high-throughput applications, \acp{PC} are typically decoded with \ac{iBDD} where the component code is decoded by an efficient algebraic component code decoder based on the hard-decision channel output. \ac{iBDD} is also often referred to as \ac{HDD} of \acp{PC}.

\Ac{SDD} of \acp{PC}, also known as \ac{TPD}~\cite{pyndiah1998near} improves the error-correcting ability of \acp{PC} by exploiting soft channel information and list-based decoding. Typically, a 1-2\,dB coding gain improvement can be observed compared to \ac{HDD}/\ac{iBDD}~\cite{ViasatTPC66100}. However, the high internal decoder data flow required by the soft-message passing in \ac{TPD} makes it challenging to adapt for ultra-high-speed optical fiber communication systems operating at throughputs of $800$ Gbit/s and beyond~\cite{sun2020800g}. In contrast, \ac{HDD} provides a significant reduction in internal decoder data flow by only passing hard messages. Recently, several \emph{hybrid} SDD/HDD schemes have been proposed which provide a performance/complexity trade-off. The unifying idea of these algorithms is to use soft channel information to aid the hard-decision decoder while keeping the complexity similarly low as in \ac{iBDD}.

One promising approach for hybrid SDD/HDD is to use ternary messages and \ac{EaE} decoding. \Ac{EaE} decoding with a stall pattern analysis was studied in~\cite{soma2021errors}, assuming miscorrection-free decoding. In~\cite{rapp2021error}, a thorough analysis of \ac{iEaED} with \ac{EMP} based on \ac{DE} takes miscorrections into account. The results show that \ac{iEaED} without miscorrection detection yields only small coding gains compared to \ac{HDD}. The \ac{BEE-PC} proposed in~\cite{sheikh2021novel} uses EaE decoding with a relatively high-cost miscorrection control and yields the best performance of hybrid SDD/HDD PC decoding so far. \ac{BEE-PC} is also based on the idea of combining properly scaled soft channel reliability with the component code decoding decision, which was proposed and developed in~\cite{sheikh2018iterative, sheikh2019binary,sheikh2021refined,sheikh2018low, sheikh2019BMPGMDD}.

\Ac{SABM} decoder was proposed in~\cite{lei2019improved} and later improved to \ac{SABM-SR} in~\cite{liga2019novel} for \acp{PC}. The bits with high channel \acp{LLR} are marked as highly reliable bits (HRBs) and used for miscorrection detection based on the principle that a BDD output is considered as miscorrection if it conflicts with any HRBs. However, the effect of erroneous HRBs are difficult to be eliminated without an effective update mechanism for the HRBs.

In~\cite{hager2018approaching}, \ac{AD} was proposed. Unlike the above-mentioned hybrid decoders, AD requires no soft channel output but instead operates purely on the hard-decision channel output. The anchor bits are dynamically set during decoding and used for miscorrection detection as in SABM. However, the achievable coding gain is limited due to wrongly marked anchor bits, especially in the first decoding iterations. AD has also been shown to be useful in reducing the decoding complexity and improving the decoding performance when combined with list decoding for low rate PCs~\cite{Senger2019Improved,Senger2019List}.

In this paper, we propose a novel hybrid decoding scheme for \acp{PC}. We first propose a modified \ac{EaE} decoder able to decode beyond the minimum distance of the component codes, at the cost of an increased miscorrection rate. The miscorrection problem is then solved with a novel miscorrection detection scheme which resembles a combination of AD and SABM. We introduce a new reliability measurement called \ac{DRS} that is initialized with the soft channel output and updated during iterative decoding. The DRSs are then used to identify the anchor bits and to detect miscorrections. We present the decoder and a detailed analysis of the decoding behavior.

The remainder of the paper is organized as follows. In Sec.~\ref{sec:pre}, the preliminaries are given. In Sec.~\ref{sec:eaed}, we introduce the proposed \ac{EaE} decoder used as the component code decoder for PCs and calculate the decoding ability of such decoder assuming no miscorrections. In Sec.~\ref{sec:algo}, we introduce the \ac{DRS} and describe the architecture of the proposed decoding algorithm. The simulation results in Sec.~\ref{sec:simu} shows that the proposed decoder yields improved decoding performance and approaches miscorrection-free decoding. The reason of the decoding performance gain is heuristically illustrated in Sec.~\ref{sec:analy} with an example. In Sec.~\ref{sec:complexity}, we analyze the computational and storage overhead. The last section concludes the paper.

\textit{Notation}: We use boldface letters to denote vectors and matrices, e.g., $\vec{y}$ and $\vec{Y}$. The $i$-th component of vector $\vec{y}$ is denoted by $y_i$, and the element at the $i$-th row and $j$-th column of $\vec{Y}$ is denoted by $Y_{i,j}$. Let $\vec{y}_i$ be the $i$-th row of a matrix $\vec{Y}$. $\mathbb{Z}_{32}$ stands for the set $\{0,1,\ldots,31\}$. $\mathbb{R}$ stands for the set of real numbers and $\mathbb{R}_{\geq 0}$ for the set of non-negative real numbers. For $x\in \mathbb{R}$, $\lfloor x \rfloor$ is the floor function that gives the greatest integer less than or equal to $x$. We use a superscript to denote the maximum number of iterations carried out by an iterative decoder, e.g., iBDD$^{10}$.

\section{Preliminaries}
\label{sec:pre}
A \ac{PC} codeword is a two-dimensional rectangular array where every row and every column is a codeword of a component code chosen to be same $(n,k,t)$ code $\CW$\footnote{We only consider \acp{PC} with identical row and column codes in this paper. However, the proposed decoding scheme naturally extends to general \acp{PC} where the row and column code are not necessarily the same.}. We consider $\CW$ being either a $(2^{\nu}-1,k_0,t)$ binary \ac{BCH} code or its $(2^{\nu}-1,k_0-1,t)$ even-weight subcode, both able to correct $t$ errors with standard \ac{BDD}. Let $\ddesign$ be the design distance of $\CW$ ($\ddesign \leq \dmin$, with $\dmin$ the minimum Hamming distance of $\CW$) and $t=\lfloor (\ddesign-1)/2 \rfloor$. The rate of the constructed \ac{PC} is $r = {k^2}/{n^2}$.

We consider a \ac{BPSK} modulation and assume that the codewords are transmitted over a \ac{BI-AWGN} channel. For any transmitted bit $x_i$, the channel output is \begin{equation*}
    \tilde{y}_i=(-1)^{x_i}+n_i,
\end{equation*} where $n_i$ is (real-valued) AWGN with noise variance $\sigma^2_n = (2r\Eb/\No)^{-1}$.

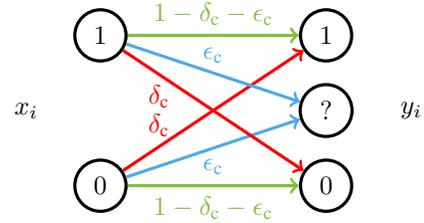
\begin{figure}[tb]	
	\centering
	\begin{tikzpicture}
\node[circle,draw=black, inner sep=4pt,very thick] (x0) at (0,0) {$0$};%
\node[circle,draw=black, inner sep=4pt,very thick] (x1) at (0,2) {$1$};%
\node(tmp1) [left=0.5cm of x0] {};%
\node(tmp2) [left=0.5cm of x1] {};%
\node (y) at ($(tmp1)!0.5!(tmp2)$) {$x_i$};%

\node[circle,draw=black, inner sep=4pt,very thick] (y0) at (3,0) {$0$};%
\node[circle,draw=black, inner sep=4pt,very thick] (y1) at (3,1) {$\que$};%
\node[circle,draw=black, inner sep=4pt,very thick] (y2) at (3,2) {$1$};%
\node (y)  [right=0.5cm of y1]{$y_i$};%

\draw[->,KITpalegreen,very thick] (x0)--node[below]{$1-\deltaC-\epsilonC$}(y0);
\draw[->,KITcyanblue,very thick] (x0)--node[below]{$\epsilonC$}(y1);
\draw[->,red,very thick] (x0)--node[above, pos=0.2]{$\deltaC$}(y2);
\draw[->,red,very thick] (x1)--node[below, pos=0.2]{$\deltaC$}(y0);
\draw[->,KITcyanblue,very thick] (x1)--node[above]{$\epsilonC$}(y1);
\draw[->,KITpalegreen,very thick] (x1)--node[above]{$1-\deltaC-\epsilonC$}(y2);
\end{tikzpicture}
	\caption{Channel model of the \ac{EaE} channel}
	\label{fig:channel}
\end{figure}
Let $\Lc \coloneqq 2/\sigma^2_n$. The channel \ac{LLR} for the \ac{BI-AWGN} channel is given by
\begin{equation*}
    L(\tilde{Y}=\tilde{y}|X)=\ln\left(\frac{\frac{1}{\sqrt{2\pi \sigma_n}}\exp\left(\frac{-(\tilde{y}-1)^2}{2\sigma_n^2}\right)}{\frac{1}{\sqrt{2\pi \sigma_n}}\exp\left(\frac{-(\tilde{y}+1)^2}{2\sigma_n^2}\right)}\right)=\Lc \cdot \tilde{y},
\end{equation*}
proportional to the magnitude of the channel output. Therefore, when performing a hard-decision at the channel output, a received value $\tilde{y}_i$ with small magnitude is considered to be unreliable. We define $T$, a configurable threshold, such that values $\tilde{y}_i\in [-T,+T]$ are declared as erasures ``$\que$''. For $|\tilde{y}_i|>T$, $y_i=\textnormal{sign}(\tilde{y}_i)$ by the usual \ac{HDD} rule. Therefore, the channel outputs are mapped to three discrete values as depicted in Fig.~\ref{fig:channel} where 
\begin{equation*}
    \operatorname{\deltaC}= \QFunc\left(\sqrt{2r\frac{\Eb}{\No}} (T + 1)\right)
\end{equation*}
is the error probability and
\begin{equation*}
    \operatorname{\epsilonC} =1 -  \QFunc\left(\sqrt{2r\frac{\Eb}{\No}} (T - 1)\right) - \QFunc\left(\sqrt{2r\frac{\Eb}{\No}} (T + 1)\right)
\end{equation*}is the erasure probability. 
When $T=0$, the \ac{EaE} channel reduces to a \ac{BSC}.

An error-only \ac{BDD} succeeds when its input word is in a Hamming sphere 
\begin{equation*}
     \mathcal{S}_t(\vec{c}) = \{\vec{y}\in \{0,1\}^n :\dH(\vec{y},\vec{c})\leq t\}
 \end{equation*}
 of radius $t$ around a codeword $\vec{c}\in \mathcal{C}$ where $\dH(\vec{y},\vec{c})$ is the Hamming distance between $\vec{y}$ and $\vec{c}$.
 
 We define the \ac{BDD} decoding rule for a binary vector $\vec{y}\in \ZO^n$ as
\begin{equation*}
    \textnormal{BDD}(\vec{y})=\begin{cases}
    \vec{c}& \exists \vec{c}\in \CW \textnormal{ such that }\vec{y}\in \Sp_t(\vec{c})\\
    \vec{y}&\textnormal{otherwise}.
    \end{cases}
\end{equation*}

Similarly, we define
\begin{equation*}
\Sp^3_t(\vec{c}) \coloneqq  \{\vec{y} \in \ZQO^n : 2 \dnE{\vec{y}}(\vec{y}, \vec{c}) + \E(\vec{y}) < \ddesign\},
 \end{equation*}as the Hamming sphere in $\ZQO^n$ for a codeword $\vec{c} \in \CW$ where $\E(\vec{y}) :=|\{i : y_i = \que \}|$ is the number of erasures of $\vec{y}$ and $\dnE{\vec{y}}(\vec{y}, \vec{c})$ is the Hamming distance between $\vec{y}$ and $\vec{c}$ at the unerased coordinates of $\vec{y}$.
 
PCs are conventionally decoded with an iterative decoding scheme where the rows and columns of the PC block are alternately decoded with the component code decoder $\DFC$ until the maximum number of iterations $L$ is reached. In the conventional \ac{iBDD} decoding scheme described in Algorithm~\ref{alg:iBDD} $\DFC$ is a \ac{BDD} decoder. In this paper, we replace $\DFC$ by an EaE decoder described later in Sec.~\ref{sec:eaed}.

Throughout this paper, we always let the result of a component decoder $\DFC$ be
\begin{equation*}
	\vec{w}\coloneqq \DFC(\vec{y})\in \CW \cup \vec{y}
\end{equation*}
where $\vec{w}=\vec{y}$ in case of a decoding failure.

A \emph{miscorrection} of $\vec{y}$ happens when $\DFC(\vec{y})=\vec{c}\in \CW$ but $\vec{c}\neq \vec{x}$, the transmitted codeword. For component BCH or RS codes with small $t$, which are often used in fiber optical communication systems, miscorrections severely degrade the decoding performance of PCs. In such systems, miscorrections are frequent and occur approximately with probability $1/t!$~\cite{mceliece1986decoder, justesen2010performance}.

\begin{algorithm}[t]
	\footnotesize
	\DontPrintSemicolon
\caption{Iterative BDD (iBDD) of PCs}\label{alg:iBDD}
\textbf{Input}: $\vec{Y}\in\ZO^{n\times n}$\;
\For(\tcp*[f]{$L$:number of iterations}){$\ell=1,2,\ldots,L$}{
\For{$i=1,2$}{
\For{$j=1,2,\ldots,n$}{
$\vec{w}_j\gets \BDD(\vec{y}_j)$
}
$\vec{Y}\gets \vec{W}^T$
}
\If{$\vec{W}$ \textnormal{is valid codeword}}{\textbf{return} $\vec{W}$}
}

\textbf{Output}: $\vec{W}\in\ZO^{n\times n}$\;
\end{algorithm}

\section{Error-and-erasure Decoding}
\label{sec:eaed}

\subsection{Error-and-erasure Decoder (EaED)}
\begin{algorithm}[t]
	\footnotesize
	\DontPrintSemicolon
\caption{\Ac{EaED}}\label{alg:eaed}
\textbf{Input}: $\vec{y}\in\ZQO^n$\;
    $E\gets$ $\E(\vec{y})$\tcp*[r]{Number of erasures in $\vec{y}$}
\lIf(\tcp*[f]{Failure}){$E\geq \ddesign$} {$\vec{w}=\yimp$}
\Else{
        $\pone,\ptwo\gets$ two random, complementary vectors in $\ZO^{E}$\;
        $\yone$,$\ytwo \in \ZO^n \gets$ $\yimp$ with erasures replaced by $\pone,\ptwo$\;
        \For{$i=1,2$}
            {$\wi \gets \BDD(\vec{y}_i)$\;
            \lIf{$\BDD(\yi)\in \CW$}{$\dH_i\gets \dnE{\vec{y}}(\vec{y}, \wi)$}
            }
        \lIf(\tcp*[f]{Failure}){$\wone\not\in \CW$ \textnormal{and} $\wtwo\not\in \CW$}{$\vec{w}\gets \vec{y}$}
        \lElseIf{$\wi\in \CW $ \textnormal{and} $\vec{w}^{(j)} \not \in \CW\;(i,j\in\{1,2\},i\neq j)$}{$\vec{w}\gets \wi$}
        \Else{
            \lIf{$d_1>d_2$}{$\vec{w}=\wtwo$}
            \lElseIf{$d_2>d_1$}{$\vec{w}=\wone$}
            \lElse{$\vec{w}\gets$ random choice from $\{\wone,\wtwo\}$}
        }
    }
\textbf{Output}: $\vec{w}\in \CW \cup \vec{y}$\;
\end{algorithm}

In this paper, we propose the following \ac{EaED}, which is a modification of~\cite[Sec. 3.8.1]{MoonBook}. Let $\vec{y} \in \ZQO^n$ be the received row/column vector, and let $\vec{w} \coloneqq  \textnormal{EaED}(\vec{y})$ be the decoding result.

If $\E(\vec{y}) \geq \ddesign$, the EaED does not decode and declares a failure, returning $\vec{w}=\vec{y}$, as a large number of erasures cannot be handled by the decoder. 

If $\E(\vec{y})< \ddesign$, the erasure positions of $\vec{y}$ are first filled with two complementary \emph{random} vectors $\pone, \ptwo \in \ZO^{\E(\vec{y})}$, i.e., $\pone+\ptwo=(1,1,\ldots,1)$, resulting in two words $\yone$,$\ytwo \in \ZO^n$. The pair of vectors $\left(\pone,\ptwo\right)$ is called a \emph{filling pattern} for the erasures. Note that $\pone$ and $\ptwo$ are not constant but generated randomly in every execution of the EaED.

Then, two \ac{BDD} steps are performed. Let 
\begin{equation*}
    \wi \coloneqq \BDD\left(\yi\right)
\end{equation*}
for $i\in\{1,2\}$. The EaED output is determined based on the two BDD outputs using the following rules:

\textbf{Case 1:} If both \ac{BDD} steps fail, set $\vec{w}=\vec{y}$. 

\textbf{Case 2:} If $\wi\in \CW$ for exactly one $\wi$, set $\vec{w}=\wi$. 

\textbf{Case 3:} If both BDD steps succeed, let
\begin{equation*}
    d_i=\dnE{\vec{y}}\left(\vec{y}, \wi\right)
\end{equation*}
for $i\in\{1,2\}$. We chose $\vec{w}=\wone$ if $d_1<d_2$ and $\vec{w}=\wtwo$ if $d_1>d_2$;  If $d_1=d_2$, one of the codewords $\wi$ is chosen at random. 

The EaED algorithm is summarized in Algorithm~\ref{alg:eaed}. Note that the EaED reduces to a conventional BDD decoder when there are no erasures, i.e., $\E(\vec{y})=0$.

Another commonly-used EaE decoder is a one-step EaE decoding algorithm\footnote{This decoder was previously referred to as EaED+ in~\cite{rapp2021error}. We do not use this name to avoid confusion with DRSD+ proposed in the later sections.} proposed by Forney in~\cite{forney1965decoding}. It extends the Gorenstein-Zierler algorithm~\cite[Ch.~6]{rothbook} and resolves errors and erasures at the same time by solving the key equation. It requires some modifications in the key equation solver inside the decoder while EaED uses two legacy BDDs for the BSC with some additional operations and control logic. Let $\vec{w}$ denote the decoding result. Forney's EaE decoder follows the decoding rule given by
\begin{equation*}
    \vec{w}=\begin{cases}
    \vec{c}& \exists \vec{c}\in \CW \textnormal{ such that } \vec{y}\in \Sp_t^3(\vec{c})\\
    \vec{y}&\textnormal{otherwise}.
    \end{cases}
\end{equation*}

We do not use Forney's EaE decoder in this paper because an \ac{EaED}, like Forney's EaE decoder, can correct any joint \ac{EaE} pattern if $\vec{y}\in \Sp^3_t(\vec{c})$~\cite[Theorem 1]{rapp2021error}.

Moreover, the \ac{EaED} may correct some \ac{EaE} patterns for $2\dnE{\vec{y}}(\vec{y}, \vec{c}) + \E(\vec{y})\geq \ddesign$ because all (or large enough number) of the erasures may possibly be filled with a correct transmit value when generating $\yone$ and $\ytwo$. Thus, the \ac{EaED} has potentially higher error-correcting capabilities than the one-step EaE decoder (see also Sec.~\ref{subsec:ideal_eaed}) but is also more prone to miscorrections without the constraint that $\vec{w}\in \Sp^3_t(\vec{y})$. Consequently, in our previous work, the EaED did not yield a satisfying decoding performance gain for \acp{PC} due to the lack of miscorrection control~\cite{rapp2021error}. In this paper, we deal with the miscorrection problem to fully exploit the error-correcting potential of the \ac{EaED}.

\subsection{Failure Analysis of EaED with Ideal Miscorrection Detection}
\label{subsec:ideal_eaed}

In this section, we analyze the error-correcting ability of a genie-aided \emph{ideal} EaED following the decoding rule given by
\begin{equation*}
    \textnormal{ideal EaED}(\vec{y})=\begin{cases}
    \vec{x}& \textnormal{if EaED}(\vec{y})=\vec{x}\\
    \vec{y}&\textnormal{otherwise},
    \end{cases}
\end{equation*}
where $\vec{x}$ is the transmit codeword. It can be seen as an EaED followed by an ideal miscorrection detection that discards all miscorrections. Although such a decoder may be impossible to realize, our simulation results in Sec.~\ref{sec:analy} show that our proposed novel decoder approaches its performance when the number of erasures is moderate.

If no miscorrections happen, the success rate of one component code decoding by ideal EaED can be calculated combinatorically. Let $\Ps(D,E)$ denote the ideal EaED success probability when decoding with $D$ errors and $E$ erasures.

\textbf{Case 1:} If $D>t$ or $E\geq \ddesign$, $\Ps(D,E)=0$. When $D>t$, miscorrection-free BDD decoding succeeds neither for $\yone$ nor $\ytwo$. When $E\geq \ddesign$, we do not decode (see Algorithm~\ref{alg:eaed}). 

\begin{figure}[tb]	
	\centering
	\includegraphics{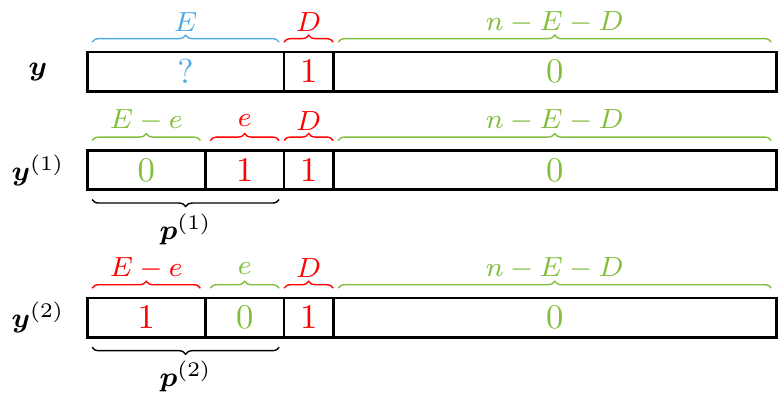}
	\caption{Graphical illustration of the number of EaEs with a filling pattern $(\vec{p}^{(1)},\vec{p}^{(2)})$ assuming the zero codeword was transmitted.}
	\label{fig:fillingpattern}
\end{figure}

\textbf{Case 2:} Assume $D\leq t,E< \ddesign$, the ideal EaED succeeds if and only if 
\begin{equation*}
    \exists i\in \{1,2\}, \BDD\left( \yi \right) = \wi = \vec{x}.
\end{equation*}
Let $e$ denote the number of positions in $\pone$ that differ from the transmit codeword $\vec{x}$. Fig.~\ref{fig:fillingpattern} illustrates the received word $\vec{y}$ and both $\yone$ and $\ytwo$ where the erasures have been replaced by the filling patterns. Under the assumption that $\vec{x}=\vec{0}$ was transmitted, $e$ denotes the number of ``$1$''s in $\pone$. Then, the number of such positions in $\ptwo$ is $E-e$ because $\pone$ and $\ptwo$ are complementary. The number of errors in $\yone$ is $D+e$ and the number of errors in $\ytwo$ is $D+E-e$. Therefore, the condition for correctability reduces to $D+e\leq t$ or $D+E-e\leq t$, i.e.,
\begin{equation}
\label{eq:irange}
    e\in [0,t-D]\cup[E+D-t,E].
\end{equation}
\textbf{Case 2.1:} If $2D+E<\ddesign\leq 2t+2$, $\Ps(D,E) = 1$ (with $\ddesign=2t+1$ for BCH codes and $\ddesign=2t+2$ for their even-weight subcodes). Assume that \eqref{eq:irange} does not hold, i.e., $t-D<e<E+D-t$ for some $e$, which is the same as saying $t-D+1\leq e\leq E-(t-D)-1$. This means that $2D+E\geq 2t+2$, contradicting $2D+E<\ddesign$.

\textbf{Case 2.2:} If $2D+E\geq \ddesign$, we calculate $\Ps(D,E)$ by counting the number of cases where \eqref{eq:irange} holds. We first show that it is impossible that $e\leq t-D$ and $e\geq E+D-t$ are true at the same time. Assume that both conditions are fulfilled, then $E+D-t \leq e \leq t-D$ for some $e$. This leads to $2D+E\leq 2t$. As $\ddesign>2t$, this is impossible. Therefore, either $\yone$ or $\ytwo$ leads to the correct transmit codeword. If $e\leq t-D$, there are $\sum_{e=0}^{t-D}\binom{E}{e}$ possible filling patterns. If $e\geq E+D-t$, there are $\sum_{e=E-\left(t-D\right)}^{E}\binom{E}{e}$ possible filling patterns. Therefore, there are
\begin{equation*}
    \sum_{e=0}^{t-D}\binom{E}{e} + \sum_{e=E-\left(t-D\right)}^{E}\binom{E}{e}= 2\sum_{e=0}^{t-D}\binom{E}{e}
\end{equation*} cases where decoding will succeed. In total, there are $2^E$ possible filling patterns for the $E$ erasures and each pattern is equally likely. Thus, $\Ps(D,E)= 2\sum_{e=0}^{t-D}\binom{E}{e}/2^E$ in this case.

Therefore, we summarize
\begin{equation*}
\Ps(D,E) =\begin{cases}
1 & 2D+E<\ddesign,\\
0&E \geq \ddesign \;\textnormal{or}\; D>t,\\
2^{1-E}\sum_{e=0}^{t-D}\binom{E}{e}&\text{otherwise}.\\
\end{cases}
\end{equation*}

\begin{table}[tb]
\caption{EaED success rate for different \ac{EaE} pattern for a $t=2,\ddesign=6$ component code}
	\begin{tabular}{c|cccccc}
		\backslashbox{$D$}{$E$}&
		0 &1 &2 &3 &4 &5\\ \hline
		0 &1.000 &1.000 &1.000 &1.000&1.000 &\multicolumn{1}{c|}{1.000} \\ 
		\cline{6-7}
		1 &1.000 &1.000 &1.000 &\multicolumn{1}{c|}{1.000} &0.625&0.375\\
		\cline{4-5}
		2 &1.000 &\multicolumn{1}{c|}{1.000} &0.500 &0.250 &0.125&0.063
	\end{tabular}
\label{tab:success_rate_even}
\vspace{2ex}
\caption{Forney's EaE decoder success rate for different \ac{EaE} pattern for a $t=2,\ddesign=6$ component code}
\begin{tabular}{c|cccccc}
		\backslashbox{$D$}{$E$}&
		0 &1 &2 &3 &4 &5\\ \hline
		0 &1.000 &1.000 &1.000 &1.000&1.000 &\multicolumn{1}{c|}{1.000} \\ 
		\cline{6-7}
		1 &1.000 &1.000 &1.000 &\multicolumn{1}{c|}{1.000} &0&0\\
		\cline{4-5}
		2 &1.000 &\multicolumn{1}{c|}{1.000} &0 &0 &0&0
	\end{tabular}
\label{tab:success_rate_evenEaED+}
\vspace{2ex}
\caption{EaED success rate for different \ac{EaE} pattern for a $t=2,\ddesign=6$ component code in $\ell=5$ decoding trials}
\begin{tabular}{c|cccccc}
		\backslashbox{$D$}{$E$}&
		0 &
		1 &
		2 &
		3 &
		4 &
		5
		\\ \hline
		0 &
		1.000 &
		1.000 &
		1.000 &
		1.000 &
		1.000 &
		\multicolumn{1}{c|}{1.000}
		\\ \cline{6-7} 
		1 &
		1.000 &
		1.000 &
		1.000 &
		\multicolumn{1}{c|}{1.000} &
		0.993 &
		0.905
		\\ \cline{4-5}
		2 &
		1.000 &
		\multicolumn{1}{c|}{1.000} &
		0.969 &
		0.762 &
		0.487 &
		0.276
	\end{tabular}
\label{tab:success_rate_even_ite}
\vspace{-2ex}
\end{table}
For example, Tab.~\ref{tab:success_rate_even} gives the success rate of EaED when decoding with various numbers of \ac{EaE}s for a $t=2,\ddesign=6$ component code. When $2D+E\geq \ddesign$, the success probability decreases with an increasing number of EaEs. However, the values are not zero as for the one-step EaE decoder by Forney (Tab.~\ref{tab:success_rate_evenEaED+}). The advantage of ideal EaED becomes larger during iterative decoding, where new, independent filling patterns are chosen each time a codeword is decoded. With ideal miscorrection detection, if the success probability of one EaED step is $\Ps$, the probability of success after $\ell$ decoding trials is $1-(1-\Ps)^{\ell}$. For example, Tab.~\ref{tab:success_rate_even_ite} shows the success probability after $\ell=5$ decoding trials for EaED with ideal miscorrection detection. This also proves that random filling patterns $(\pone,\ptwo)$ are crucial for the performance of the proposed decoder. As the number of \ac{EaE}s is reduced during the \ac{PC} decoding process, the actual success rate can be potentially even higher. Moreover, the decoding success probability $1-(1-\Ps)^{\ell}$ approaches $1$ with a sufficiently large number of iterations, which is not realistic in practice. Thus, we only use 10 decoding iterations of the ideal EaED when using it as a benchmark.

\begin{figure*}[tb]	
	\centering
	\includegraphics[width = \textwidth]{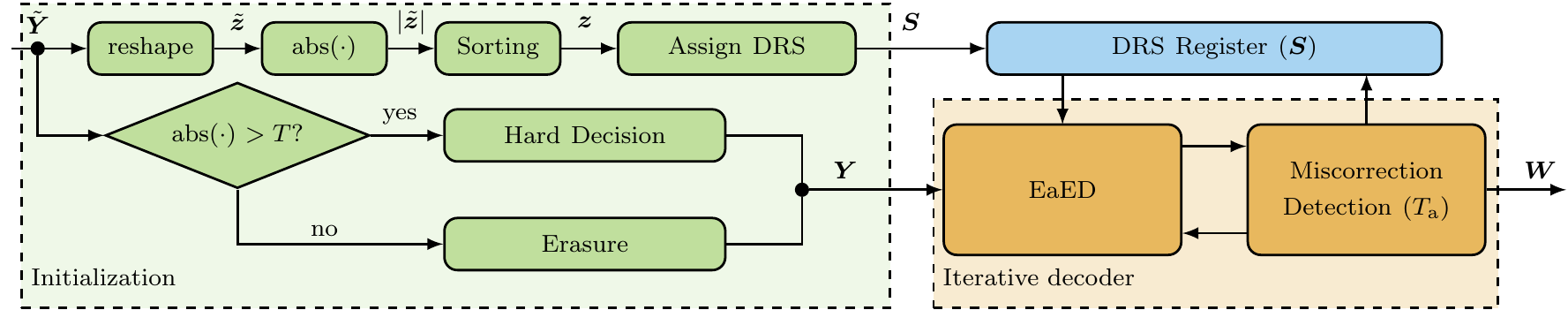}
	\caption{Block diagram of the proposed DRSD.}
	\label{fig:blockDiagram}
\end{figure*}

\section{Dynamic Reliability Score (DRS) Decoding}
\label{sec:algo}

To represent and update the reliability of a bit in the PC block, we introduce a reliability measure called \ac{DRS}, which is stored in an additional register separately from the PC codeword. The \ac{DRS} reflects the reliability of a bit from both its channel \ac{LLR} and its behavior during decoding. We propose to represent the DRSs by 5-bit integers in the range $[0,31]$ (as we observe that further increasing the number of representation levels does not improve the performance of the proposed decoder significantly). We introduce an anchor threshold $\Ta$. All bits with a \ac{DRS}~$> \Ta$ are classified as \emph{anchor~bits} (following~\cite{hager2018approaching}) during decoding and are not allowed to be flipped by a component code decoder. Note that the \ac{DRS} is not a fully-accurate measurement of the reliability, i.e., we cannot say that a bit with higher DRS must be more reliable than a bit with lower DRS. However, we can say that bits with DRS above the threshold $\Ta$ can be considered as correct with high confidence. The DRS is used for miscorrection detection in iterative decoding with EaED.

The block diagram and workflow of the proposed 
\ac{DRSD} is shown in Fig.~\ref{fig:blockDiagram}.

\begin{algorithm}[t]
	\footnotesize
	\DontPrintSemicolon
\caption{Initialization of the Dynamic Reliability Scores (DRSs)}\label{alg:DRSDini}
\textbf{Input}: $\vec{\tilde{Y}}\in \mathbb{R}^{n\times n}$\;
$\tilde{\vec{z}}\gets$ reshape ($\tilde{Y}$) \tcp*[r]{Eq. \eqref{eq:iniDRS1}}
$\vec{z} \gets $ sort $|\tilde{\vec{z}}|$ ascendingly\;
$\sigma(\cdot) \gets$ permutation function of the indices\;
$\vec{S}\gets $ assign DRS according to $\vec{z}$\tcp*[r]{Eq. \eqref{eq:iniDRS2}}
\textbf{Output}: $\vec{S}\in \{9,10,\ldots,24\}^{n\times n}$\;
\end{algorithm}

\begin{algorithm}[t]
	\footnotesize
	\DontPrintSemicolon
\caption{Dynamic Reliability Score Decoder (DRSD) and the DRSD+ termination option}\label{alg:DRSD}
\textbf{Input}: $\vec{\tilde{Y}}\in \mathbb{R}^{n\times n}$\;
$\vec{Y}\gets \vec{0}^{n\times n}$\;
\For(\tcp*[f]{Initialization}){$i=1,2,\ldots,n$}{
\For{$j=1,2,\ldots,n$}{
\lIf{$|\tilde{Y}_{i,j}|\leq T$}{$\vec{y}_{i,j}=\que$}
\lElseIf{$\tilde{Y}_{i,j}>T$}{$\vec{y}_{i,j}=0$}
\lElse{$\vec{y}_{i,j}=1$}
}
}
$\vec{S}\gets$ initial DRSs\tcp*[r]{Algorithm \ref{alg:DRSDini}}
\For(\tcp*[f]{Decoding with DRS}){$\ell=1,2,\ldots,L-L / 5$}{
\For{$i=1,2$}{
\For{$j=1,2,\ldots,n$}{
\If{$\textnormal{Syndrome}(\vec{y}_j)=\vec{0}$}{\For{$m\in\{1,2,\ldots,n\}$}{$S_{j,m}=\min(S_{j,m}+1,31)$\;
} \textbf{continue}}
$\vec{w}_{\textnormal{tmp}}\gets \EaED(\vec{y}_j)$\tcp*[r]{Algorithm~\ref{alg:eaed}}
$\vec{k}\gets \vec{0}^{n}$\;
\For{$m=1,2,\ldots,n$}
{
\lIf{$Y_{j,m}\neq \que$}
{$k_{m}=w_{\textnormal{tmp},m}\oplus Y_{j,m}$}
}
$\vec{a}\in \ZO^{n}$, $a_m=\ind{S_{j,m}>\Ta}$\;
\If(\tcp*[f]{Not miscorrection}){\textnormal{sum}$(\vec{k}\cdot \vec{a}^{T})=0$}{$\vec{w}_j\gets \vec{w}_{\textnormal{tmp}}$\;
}

\For{$m\in\{1,2,\ldots,n\}$}{\If{$k_{m}$}{$S_{j,m}=\max(S_{j,m}-1,0)$}
}
}

$\vec{Y}\gets \vec{W}^T$, $\vec{S}\gets\vec{S}^{T}$\;
\lIf{$\vec{W}$\textnormal{is valid codeword}}{\textbf{return} $\vec{W}$}
}
\If(\tcp*[f]{Increase anchor threshold}){$\ell \textnormal{ mod } 5=0$}{$\Ta=\Ta+1$}
}

\For(\tcp*[f]{Termination}){$\ell=1,2,\ldots,L / 5$}{
    \If {\text{DRSD}}{
        \For(\tcp*[f]{Plain iEaED}){$i=1,2$}
        {   
            \For{$j=1,2,\ldots,n$}{
                $\vec{w}_j\gets \EaED(\vec{y}_j)$\tcp*[f]{Algorithm~\ref{alg:eaed}}
            }
        $\vec{Y}\gets \vec{W}^T$\;
        }
        \lIf{$\vec{W}$ \textnormal{is valid codeword}}{\textbf{return} $\vec{W}$}
    }
    \If {\text{DRSD+}}{
        repeat line 10-27 with $\Ta=\Ta^*$
        }
    }

Replace erasures in $\vec{W}$ with random binary value\;
\textbf{Output}: $\vec{W}\in\ZO^{n\times n}$\;
\end{algorithm}

\begin{figure*}[tb]	
\centering
	\vspace{-1ex}
	\includegraphics{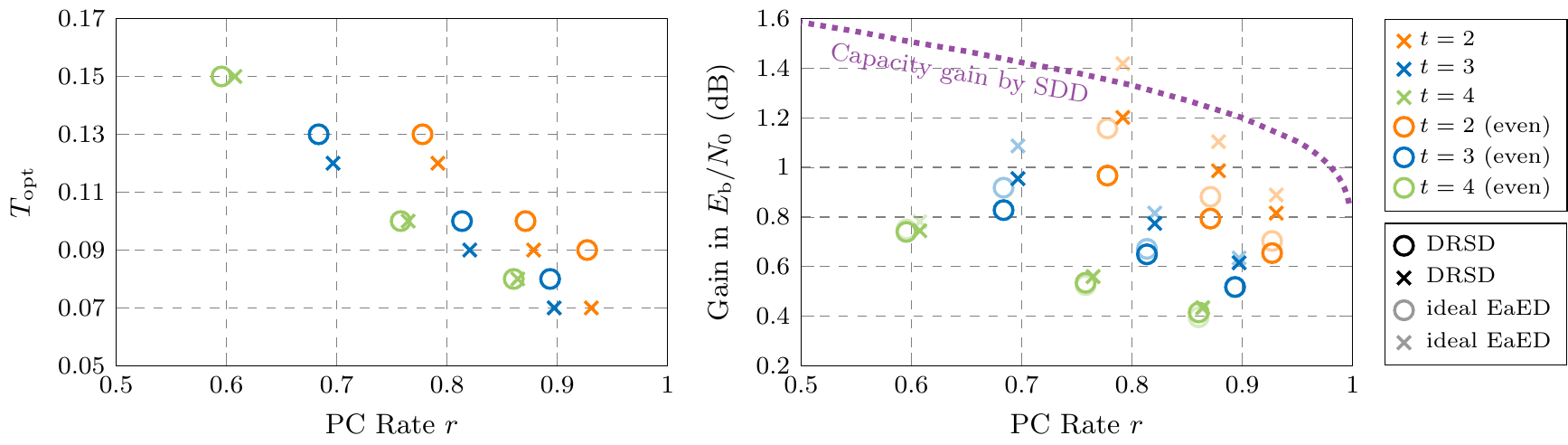}
	\vspace{-1ex}
	\caption{Simulation results of the parameter analysis: The optimal erasure threshold $T_{\textnormal{opt}}$ for \ac{DRSD}$^{20}$ found during numerical simulation is plotted in the left subplot. The noise threshold gain that the \ac{DRSD}$^{20}$ achieves compared to \ac{iBDD}$^{20}$ is plotted in the right subplot with the non-transparent markers. The component codes are $(n,k,t)-$\ac{BCH} codes with $n\in \{127,255,511\}$ and $t\in \{2,3,4\}$ or their even-weight subcodes. The abscissa corresponds to the rate $r$ of the constructed PCs. Additionally on the right subplot, the transparent markers corresponds to the respective EaED with ideal miscorrection detection and $L=16$ iterations (denoted as ``ideal EaED''). The dotted curve marks the capacity gain due to \ac{SDD} on the binary-input AWGN channel.}
	
	\label{fig:gain}
\end{figure*}
At the initialization, the received unquantized real-valued channel output of the \ac{PC} block $\tilde{\vec{Y}}\in \mathbb{R}^{n\times n}$ is fed to two paths. In the upper path, the initial value of the DRSs $\vec{S}$ is calculated for the PC block. To do this, we first reshape $\tilde{\vec{Y}}$ into a vector $\tilde{\vec{z}}=(\tilde{z}_1,\tilde{z}_2,\ldots,\tilde{z}_{n^2})\in \mathbb{R}^{n^2}$ where 
\begin{equation}
\label{eq:iniDRS1}
    \tilde{z}_{n(i-1) + j}=\tilde{Y}_{i,j}
\end{equation}
Then the absolute value of $\tilde{\vec{z}}$ is taken and sorted ascendingly, resulting in a vector $\vec{z}=(z_1,z_2,\ldots,z_{n^2})\in \mathbb{R}^{n^2}_{\geq 0}$ with
\begin{equation*}
    z_i\geq z_j,\forall i>j,
\end{equation*} 
where
\begin{equation*}
    z_{\sigma(i)}=|\tilde{z}_i|
\end{equation*}
and $\sigma(\cdot)$ is a permutation of the indices corresponding to the sorting operation. Then, the initial DRSs $\vec{S}\in \mathbb{Z}_{32}^{n\times n}$ are assigned by
\begin{equation}
\label{eq:iniDRS2}
    S_{i,j}=9+\left \lfloor  \frac{ 16\cdot (\sigma \left (n \left(i-1\right)+j \right)-1) }{n^2} \right \rfloor,
\end{equation}
such that the initial DRSs are in the range of [9,24]. This process is summarized in Algorithm~\ref{alg:DRSDini}. This initial \ac{DRS} value is stored in the \ac{DRS} register. The anchor threshold $\Ta$ is set as the optimal value found during simulation (see Sec.~\ref{sec:simu}).

In the lower branch, erasures are marked, i.e., $Y_{i,j}=\que$ if $|\tilde{Y}_{i,j}|<T$, with  $T$ (defined in Sec.~\ref{sec:pre}) being an optimizable threshold. For non-erased bits, a usual hard-decision is performed. The initial block $\vec{Y}\in \ZQO^{n\times n}$ is passed to the iterative row and column decoder.

During iterative decoding, the rows and columns are decoded alternately until the maximal number of iterations $L$ is reached, as in a conventional iBDD decoder. We replace the BDD decoder by an EaED. Moreover, every EaED decoding step is followed by the miscorrection detection step evaluating whether an anchor bit is flipped by the EaED output. In this case, this decoder output is discarded and the \ac{DRS} for all the anchor bits in conflict with this EaED output is reduced by $1$. If a decoding step does not flip any anchor bit, this EaED output will be accepted while the \ac{DRS} of all flipped bits is reduced by $1$. If a vector is already a codeword and thus no decoding is performed, the \ac{DRS} of every bit in it is increased by $1$. This is based on the observation that the longer a bit stays constant during iterative decoding, the more reliable it is. A similar idea has been exploited in AD~\cite{hager2018approaching}. The messages used for updating the DRS are all set to $1$ to enable a simple hardware implementation of the DRS register avoiding the need for full adders. The DRS values will be clipped to $0$ and $31$, respectively. In the case of an EaED decoding failure, neither the codeword nor the \ac{DRS} is changed. In addition, the anchor threshold $\Ta$ is increased by $1$ after every five decoding iterations, such that a small penalty is given for words that fail to decode consistently. With the update of the \acp{DRS}, the anchor bits are reevaluated every iteration. For \ac{DRSD}$^{L}$, the last $L/5$ iterations are performed with plain \ac{iEaED}. This is to eliminate the effect of erroneous anchor bits with high DRS. After $L$ iterations, the erased values which have not been resolved are replaced with a random binary value. We call the proposed decoder \ac{DRSD} and the complete algorithm of the proposed decoding scheme is summarized in Algorithm~\ref{alg:DRSD}.

Furthermore, when $L$ is sufficiently large, (e.g. $L=20$), we propose an alternative termination option where the few final decoding iterations are no longer plain iEaED but DRSD with a very high and constant but optimizable anchor threshold $\Ta^*$ (see Algorithm~\ref{alg:DRSD}, lines 37-38). We call the resulting algorithm DRSD+. The DRSD+ is motivated by the observation that most of the correct bits will have very high DRSs after a sufficiently large number of iterations (see Sec.~\ref{sec:analy}). Consequently, when a large $\Ta^*$ is used, the erroneous anchor bits are removed while a relatively large set of correct anchor bits are still preserved, resulting in a decoding performance improvement.

\section{Simulation Results}
\label{sec:simu}

\subsection{Preparation}

We verify and demonstrate the performance of DRS decoding using numerical examples. For the proposed \ac{DRSD}, we consider both $L=10$ and $L=20$ as $10$ iterations are often used in the literature for soft-aided PC codewords, but we observe that \ac{DRSD} can benefit from more than $10$ iterations.

We determine the \emph{noise threshold} as the minimal $\Eb/\No$ with which the target \ac{BER} of $10^{-4}$ after a fixed number of iterations is achieved. For two decoders, we compare their performance by calculating the \emph{noise threshold difference (gain)} $\Delta(\Eb/\No)^{*}$. The noise thresholds are calculated numerically by a Monte Carlo approach along with a binary search and both the optimal erasure threshold $T$ ($T_{\textnormal{opt}}$) and the optimal initial anchor threshold $\Ta$ are found during the search. As the initial DRSs are in the range of $[9,24]$, the initial $\Ta$ is also in this range. Moreover, as a too small number of anchor bits does not contribute to the miscorrection detection significantly, we narrow down the search space for $\Ta$ to $[9,15]$. Then, the parameter optimization for both $T$ and $\Ta$ can be carried out by a grid search. The anchor threshold $\Ta$ depends mostly on $t$. For $L=20$, the estimated optimal initial value of the anchor threshold is $\Ta=9$ for $t=2$, $\Ta=10$ for $t=3$, and $\Ta=12$ for $t=4$. For the $(127,k,4)$ component codes, we exceptionally find $\Ta=14$. For $L=10$, the initial anchor threshold $\Ta$ is reduced by $1$ for all component codes. The optimal erasure thresholds $\Topt$ are as shown in Fig.~\ref{fig:gain}. We observe that slightly varying $T$ (within $\pm 10\%$) does not degrade the performance severely.
\begin{figure*}[htbp]
	\centering
	\definecolor{mygreen}{rgb}{0.69, 0.87, 0.541}%
\definecolor{myblue}{rgb}{0,0.4470,0.7410}
\definecolor{myblack}{rgb}{0.2,0.2,0.2}
    \begin{tikzpicture}
       \pgfplotsset{grid style={dashed, gray}}
       \pgfplotsset{every tick label/.append style={font=\footnotesize}}
        \begin{axis}[%
        name=ax1,
        width=8.5cm,
        height=6.5cm,
        xmin=2.9,
        xmax=4.9,
        ymode=log,
        ymin=1e-7,
        ymax=0.1,
        yminorticks,
        axis background/.style={fill=white, mark size=1.5pt},
        xmajorgrids,
        ymajorgrids,
        yminorgrids,
        xtick={2.1,2.3,...,5.3},
        ytick={0.1,0.01,0.001,1e-4,1e-5,1e-6,1e-7,1e-8},
        ylabel={Post-FEC BER},
        xlabel={$\Eb/\No$ (dB)},
        label style={font=\small},
        legend cell align={left},
        legend style={anchor = north east, at={(axis cs:4.9,0.1)},draw=none, fill opacity=0.7, text opacity = 1,legend columns=1, row sep = 0pt,font=\footnotesize}
]
 \addplot [color=colorDRSD, line width=0.9pt, mark=*, mark options={solid, colorDRSD, fill=white, mark size=1.5pt}]
   table[row sep=crcr]{%
3.03649 0.0351582\\
3.11983 0.0291384\\
3.20316 0.0213643\\
3.28649 0.0135962\\
3.36983 0.0069004\\
3.45316 0.00249271\\
3.53649 0.000454642\\
3.61983 5.12845e-05\\
3.70316 3.67004e-06\\
3.78649 2.16932e-07\\
3.86983 2.13473e-08\\
 };
 \addlegendentry{DRSD$^{10}$}
 
   \addplot [color=colorDRSD, line width=0.9pt, dashed, mark=*, mark options={solid, colorDRSD, fill=white, mark size=1.5pt}]table[row sep=crcr]{%
3.03649 0.0296026\\
3.11983 0.022856\\
3.20316 0.0141033\\
3.28649 0.00503048\\
3.36983 0.00150155\\
3.45316 0.000165315\\
3.53649 2.1766e-05\\
3.61983 1.02171e-06\\
3.70316 6.83611e-08\\
 };
\addlegendentry{DRSD$^{20}$}

    \addplot [color=colorDRSDplus20, line width=0.9pt, mark=diamond*, mark options={solid, colorDRSDplus20, mark size=1.5pt}]table[row sep=crcr]{%
3.03649 0.0243675\\
3.11983 0.0160531\\
3.20316 0.00792504\\
3.28649 0.00318682\\
3.36983 0.000624824\\
3.45316 6.83286e-05\\
3.53649 3.63385e-06\\
3.61983 2.50748e-07\\
3.70316 1.33912e-08\\
 };
\addlegendentry{DRSD+$^{20}$}

 \addplot [color=myblack, line width=0.9pt, dotted, mark=square*, mark options={solid, myblack, fill=white, mark size=1pt}]
   table[row sep=crcr]{%
3.03649 0.017426\\
3.11983 0.00927323\\
3.20316 0.00373593\\
3.28649 0.000849354\\
3.36983 0.000103219\\
3.45316 9.05872e-06\\
3.53649 3.61681e-07\\
3.61983 1.25232e-08\\
};
\addlegendentry{ideal EaED$^{10}$}

\addplot [color=KITpalegreen!80, line width=0.9pt, mark=diamond*, mark options={solid,KITpalegreen!80,fill=white, mark size=1.5pt}]
  table[row sep=crcr]{%
3.44693 0.0322138\\
3.6136 0.0262655\\
3.78027 0.0220589\\
3.94693 0.015239\\
4.1136 0.00810881\\
4.28027 0.0016166\\
4.44693 0.000137922\\
4.6136 3.80668e-06\\
4.78027 1.04924e-07\\
4.93785 2.01883e-08\\
} node [pos=0.5,anchor=south,font=\footnotesize,sloped] {iBDD$^{10}$};

\addplot [color=myblue, line width=0.9pt, mark=triangle*, mark options={solid, myblue, fill=white, mark size=1.5pt}]
  table[row sep=crcr]{%
3.08263052328897	0.0435844623698019\\
3.18263052328897	0.0389537160310126\\
3.28263052328897	0.0330722844388754\\
3.38263052328897	0.0224191401049417\\
3.48263052328897	0.0139360952306367\\
3.58263052328897	0.0047877672488057\\
3.68263052328897	0.00121058814315921\\
3.78263052328897	8.37610975444792e-05\\
3.88263052328897	5.20241441221941e-06\\
3.98263052328897 1.86711340176776e-07\\
4.098263052328897 4.86711340176776e-09\\
} node [pos=0.5,anchor=south,font=\footnotesize,sloped] {SABM-SR$^{10}$~\cite{liga2019novel}};

\addplot[color=black, line width=0.9pt, mark=square*, mark options={solid, black, fill=white, mark size=1pt}]
  table[row sep=crcr]{%
2.9	0.0202655407357798\\
3	0.00509089398311894\\
3.1	0.000466080054785853\\
3.2	1.29053451418654e-05\\
3.3	4.71449588895136e-07\\
3.4	3.91569425992638e-09\\
} node [pos=0.5,anchor=south,font=\footnotesize,sloped] {TPD$^{12}$~\cite{pyndiah1998near}};
\draw[latex-latex,very thick] (axis cs:3.62,2.2e-7)--(axis cs:4.76, 2.2e-7);
\node[font=\footnotesize] at (axis cs:4.3,4e-7) {1.14\,dB};

\end{axis}
\begin{axis}[%
         at={(ax1.south east)},
        xshift=1.5cm,
        xmin=3.6,
        xmax=5.3,
        ymode=log,
        ymin=1e-7,
        ymax=0.1,
        yminorticks=true,
        axis background/.style={fill=white, mark size=1.5pt},
        xmajorgrids,
        ymajorgrids,
        yminorgrids,
        width=9cm,
        height=6.5cm,
        xtick={3.5,3.7,...,5.5},
        ytick={0.1,0.01,0.001,1e-4,1e-5,1e-6,1e-7,1e-8},
        xlabel={$\Eb/\No$ (dB)},
        ylabel={Post-FEC BER},
        label style={font=\small},
        legend cell align={left},
        legend style={anchor = north east,  at={(axis cs:5.3,0.1)}, draw=none, fill opacity=0.7, text opacity = 1,legend columns=1,font=\footnotesize, row sep = 0pt}
]
 \addplot [color=colorDRSD, line width=0.9pt, mark=*, mark options={solid, colorDRSD, fill=white, mark size=1.5pt}]
 table[row sep=crcr]{%
 	3.89162 0.0175631\\
 	3.97079 0.0135286\\
 	4.04995 0.00847037\\
 	4.12912 0.00311496\\
 	4.20829 0.00036694\\
 	4.28745 8.50604e-06\\
 	4.36662 3.0055e-08\\
 };
\addlegendentry{DRSD$^{10}$}

 \addplot [color=colorDRSD, dashed, line width=0.9pt, mark=*, mark options={solid, colorDRSD, fill=white, mark size=1.5pt}]
   table[row sep=crcr]{%
3.89162 0.0150249\\
3.97079 0.0089769\\
4.04995 0.00249982\\
4.12912 0.000224489\\
4.20829 3.30875e-06\\
4.28745 9.83661e-09\\
 };
 \addlegendentry{DRSD$^{20}$}
 
  \addplot [color=colorDRSDplus20, line width=0.9pt, mark=diamond*, mark options={solid, colorDRSDplus20, mark size=1.5pt}]
   table[row sep=crcr]{%
3.89404 0.012168\\
3.97321 0.00502814\\
4.05238 0.000890619\\
4.13154 2.5053e-05\\
4.21071 2.1675e-07\\
4.28988 4.5108e-10\\
 };
 \addlegendentry{DRSD+$^{20}$}

\addplot [color=myblack, line width=0.9pt, dotted, mark=square*, mark options={solid, myblack, fill=white, mark size=1pt}]
table[row sep=crcr]{%
3.89404 0.0101831\\
3.97321 0.00541643\\
4.05238 0.00137433\\
4.13154 7.59973e-05\\
4.21071 6.07155e-07\\
4.28988 1.39937e-09\\
};
\addlegendentry{ideal EaED$^{10}$}

 \addplot [color=colorBEE-PC, solid, line width=0.9pt, mark=pentagon*, mark options={mark size=1.5pt, colorBEE-PC,fill =white}]
 table[row sep=crcr]{%
 	3.5	0.0330750221027553\\
 	3.6	0.0308023240929515\\
 	3.7	0.0281203090160573\\
 	3.8	0.0252169175438972\\
 	3.9	0.021902817410993\\
 	4	0.0178913666084907\\
 	4.1	0.0130076314368552\\
 	4.18	0.00604362408556861\\
 	4.23	0.00248026574602935\\
 	4.28	0.000628884366079135\\
 	4.34	8.82687260395475e-05\\
 	4.39	6.04359294357079e-06\\
 	4.46	2.61493201490553e-08\\
 };
\addlegendentry{BEE-PC$^{12}$~\cite{sheikh2021novel}}

 	\addplot [color=colorAD, solid, line width=1.0pt, mark=*, mark options={mark size=1pt,solid, colorAD}]
 table[row sep=crcr]{%
 	4.92793756778681	6.65897731641676e-08\\
 	4.90068913615227	3.97347174163783e-07\\
 	4.87381045464267	1.97354863514033e-06\\
 	4.84728893195624	7.28989239655492e-06\\
 	4.82111263021807	2.46572364904559e-05\\
 	4.79527021979927	7.4733407469052e-05\\
 	4.7697509380099	0.000184188952256648\\
 	4.74454455127162	0.000426733921677219\\
 	4.71964132042203	0.000820476030078872\\
 	4.69503196884302	0.00152385841334881\\
 	4.67070765314	0.00233855221508121\\
 	4.64665993612949	0.00338809752324448\\
 	4.62288076191923	0.00454825224785916\\
 	4.59936243288797	0.00571935182468557\\
 	4.57609758839282	0.0067410803261563\\
 	4.5530791850501	0.00767428570573724\\
 } node [pos=0.4,anchor=south,font=\footnotesize,sloped] {AD$^{12}$~\cite{hager2018approaching}};;

\addplot [color=KITpalegreen!80, line width=0.9pt, mark=diamond*, mark options={solid,KITpalegreen!80,fill=white, mark size=1.5pt}]
  table[row sep=crcr]{%
4.18365 0.0168525\\
4.34198 0.0140363\\
4.50032 0.0107882\\
4.65865 0.00708795\\
4.73782 0.00423753\\
4.81698 0.00200887\\
4.89615 0.000396122\\
5.01873 7.74358e-06\\
5.09789 2.14556e-07\\
5.17706 7.43846e-09\\
} node [pos=0.5,anchor=south,font=\footnotesize,sloped] {iBDD$^{10}$};;

\addplot [color=myblue, line width=0.9pt, mark=triangle*, mark options={solid, myblue, fill=white, mark size=1.5pt}]
  table[row sep=crcr]{
3.39684128727424	0.0351557220636894\\
3.49684128727424	0.032829957458728\\
3.59684128727424	0.0312758880271704\\
3.69684128727424	0.0289893384219464\\
3.79684128727424	0.0263360235289998\\
3.89684128727424	0.0237651651756797\\
3.99684128727424	0.0194737487088811\\
4.09684128727424	0.0153251868839831\\
4.19684128727424	0.00842142119360655\\
4.29684128727424	0.00157385199838938\\
4.39684128727424	5.06474029346528e-05\\
4.49684128727424	1.78271286267877e-07\\
4.59684128727424	6.78271286267877e-10\\
} node [pos=0.4,anchor=south,font=\footnotesize,sloped] {SABM-SR$^{10}$~\cite{liga2019novel}};;

 \addplot[color=black, line width=0.9pt, mark=square*, mark options={solid, black, fill=white, mark size=1pt}]
 table[row sep=crcr]{%
 	3.725 1e-3\\
 	3.75 1e-4\\
 	4 1e-15\\
 } node [pos=0.15,anchor=south,font=\footnotesize,sloped] {TPD (Viasat, 15OH)~\cite{ViasatTPC66100}};;

\addplot[color=black, line width=0.9pt, mark=square*, mark options={solid, black, fill=white, mark size=1pt}]
table[row sep=crcr]{%
		3.2	0.054437587082942\\
		3.3	0.046984287279358\\
		3.4	0.042677157346018\\
		3.5	0.035219971639152\\
		3.6	0.030124939456009\\
		3.7	0.0213965988332\\
		3.8	0.00829490177189\\
		3.9	0.000259890294429629\\
		3.925 7.4173e-05\\
		3.955 1.3443e-05\\
		3.9745 2.4169e-06\\
		4     1.6083e-07\\	
		4.02     1e-08\\	
	} node [pos=0.5,anchor=south,font=\footnotesize,sloped] {TPD$^{12}$~\cite{pyndiah1998near}};;

\draw[very thick,latex-latex] (axis cs:4.2,2.2e-7)--(axis cs:5.10, 2.2e-7);
\node[font=\footnotesize] at (axis cs:4.7,4e-7) {0.89\,dB};

\end{axis}
\end{tikzpicture}  
	\caption{BER vs. $\Eb/\No$ (in dB) curve for product codes of rate $0.78$ (28\% overhead, component code $\CW_1$, left plot) and $0.87$ (15\% overhead, component code $\CW_2$, right plot). The superscript of the labels denotes the maximum number of decoding iterations $L$.}
	\vspace{-2ex}
	\label{fig:BERcurve}
\end{figure*}
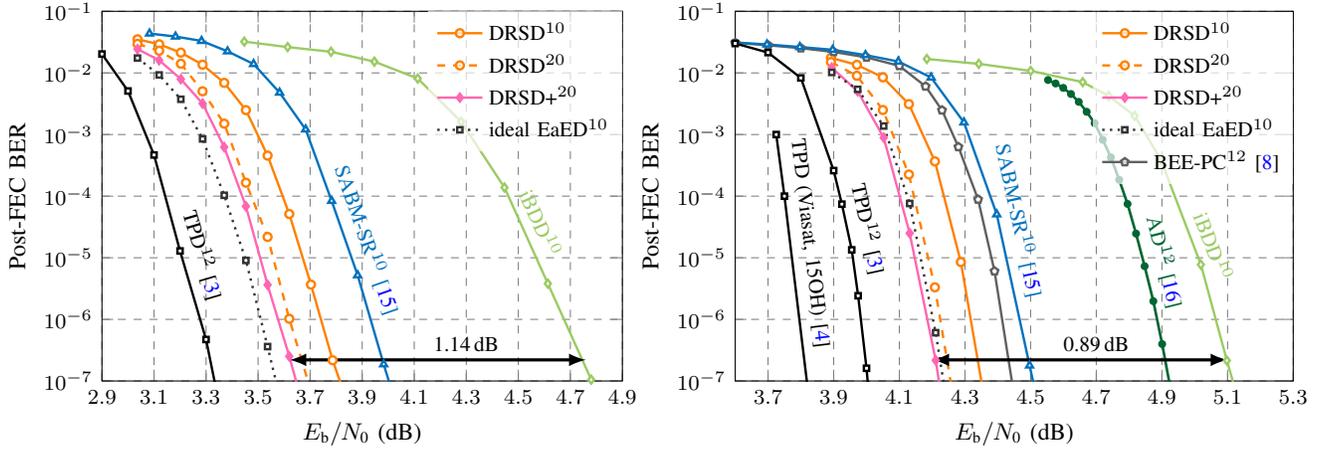

\subsection{Noise Threshold Gain $\Delta(\Eb/\No)^{*}$}

The noise threshold gain $\Delta\left(\Eb/\No\right)^{*}$ that a DRSD$^{20}$ achieves compared to an iBDD$^{20}$ decoder is calculated and shown in Fig.~\ref{fig:gain} with the non-transparent markers. We additional show the noise threshold gain of EaED with ideal miscorrection detection and $L=16$ iterations compared to iBDD$^{20}$ with the transparent markers (the optimal threshold $T$ may differ slightly but is not plotted for the sake of clarity). We use $L=16$ for EaED with ideal miscorrection detection as it does not require the final $4$ iterations of EaED to mitigate the risk of erroneous anchor bits. The dotted line depicts the capacity gain of the \ac{BI-AWGN} channel compared to the hard-decision BSC. Note that when a data point exceeds the dotted line, the code does not operate beyond capacity. This simply means that the noise threshold gain of DRSD compared to iBDD is larger than the capacity gain of the BI-AWGN channel compared to the BSC.  In general, for PCs with various component codes, the noise threshold gain of \ac{DRSD} is similarly large as the noise threshold gain of EaED with ideal miscorrection detection. This means that our proposed decoder is nearly miscorrection-free. For larger $t$ and higher code rate, the gap between \ac{DRSD} and the ideal case becomes even smaller.

\subsection{BER Results}
In Fig. \ref{fig:BERcurve}, we compare the residual post-decoding \ac{BER} for different decoders. We construct two PCs from even-weight subcodes of BCH codes denoted by $\mathcal{C}_1(127,112,2)$ and $\mathcal{C}_2 (255,238,2)$ with \ac{PC} rates of $0.78$ ($28$\% overhead) and $0.87$ ($15$\% overhead), respectively. For reference, we show the results of \ac{TPD}~\cite{pyndiah1998near}, \ac{SABM-SR}~\cite{liga2019novel}, BEE-PC~\cite{sheikh2021novel}, and \ac{AD}~\cite{hager2018approaching} with the data points provided in~\cite{sheikh2021novel} and~\cite{liga2019novel} where the component codes are singly-extended BCH codes with parameters $(128,113,2)$ and $(256,239,2)$ (hence with a small, negligible rate difference compared to our codes). For iBDD, we observe that the decoding results of $L=10$ and $L=20$ do not have a noticeable difference. Thus, we only show the results of iBDD$^{10}$. For both \acp{PC}, an obvious gain compared to \ac{iBDD} can be observed with DRSD and the performance is further improved by running $20$ iterations. With twice the number of iterations, our decoder behaves very closely to an \ac{EaED} with ideal miscorrection detection (denoted as ``ideal EaED$^{L}$'' in the figure) and has only a small gap to the significantly more complex TPD. For the rate $0.78$ \ac{PC}, we estimate, at a residual \ac{BER} of $10^{-15}$, an \ac{NCG} of $10.88$ dB (DRSD$^{10}$) and $11.03$ dB (DRSD$^{20}$). The NCGs are obtained by extrapolating the simulation results from Fig.~\ref{fig:BERcurve} using \texttt{berfit} in MATLAB${}^{\text{\textregistered}}$ assuming no error floor. For the rate $0.87$ \ac{PC}, the conjectured \acp{NCG} are $10.51$\,dB and $10.59$\,dB respectively. The modified DRSD+$^{20}$ provides an additional $0.05\;$dB coding gain compared to DRSD$^{20}$ without extra complexity. The optimal erasure threshold for DRSD+$^{20}$ is $\Topt+0.2$ with $\Topt$ the optimal erasure threshold of DRSD$^{20}$ and the optimized $\Ta^{*}=24$. Compared to a plain iBDD decoder, the DRSD+$^{20}$ yields $1.14\;$dB and $0.89\;$dB decoding gain for the rate $0.78$ and rate $0.87$ PCs, respectively. (As DRSD+$^{10}$ does not improve the decoding performance significantly, its results are not shown).

We observe that DRSD yields significantly better performance when 20 iterations are used. The reason behind this is that: first, the decoder needs to obtain a both large and reliable set of anchor bits in the first few iterations of decoding as we will show in Sec.~\ref{sec:analy}. Second, assuming no miscorrections, the use of random filling of the erasures ensures that the probability of correctly decoding words with erasures rises with the number of iterations as shown in Sec.~\ref{sec:eaed}. Moreover, as Sec.~\ref{sec:complexity} will show, using 20 iterations does not increase the decoding complexity significantly. We conclude that it is much more beneficial to use 20 decoding iterations.

\begin{figure}[tb]
	\centering
	    \begin{tikzpicture}
       \pgfplotsset{grid style={dashed, gray}}
       \pgfplotsset{every tick label/.append style={font=\footnotesize}}
       \begin{axis}[%
        xmin=2.6,
        xmax=4.5,
        ymode=log,
        ymin=1e-7,
        ymax=0.1,
        yminorticks=true,
        axis background/.style={fill=white, mark size=1.5pt},
        xmajorgrids,
        ymajorgrids,
        yminorgrids,
        width=8.5cm,
        height=6.5cm,
        xtick={2.0,2.2,...,5.6},
        ytick={0.1,0.01,0.001,1e-4,1e-5,1e-6,1e-7,1e-8},
        xlabel={$\Eb/\No$ (dB)},
        ylabel={Post-FEC BER},
        label style={font=\small},
        legend cell align={left},
        legend style={anchor = south west, at={(axis cs:2.6,1e-7)},draw=none, fill opacity=0.7, text opacity = 1,legend columns=1, row sep = 0pt,font=\footnotesize}
]
 \addplot [color=colorDRSD, line width=0.9pt, mark=*, mark options={solid, colorDRSD, fill=white, mark size=1.5pt}]
 table[row sep=crcr]{%
 	3.89162 0.0175631\\
 	3.97079 0.0135286\\
 	4.04995 0.00847037\\
 	4.12912 0.00311496\\
 	4.20829 0.00036694\\
 	4.28745 8.50604e-06\\
 	4.36662 3.0055e-08\\
 };
\addlegendentry{DRSD$^{10}$}

 \addplot [color=colorDRSD, line width=0.9pt, mark=*, dashed, mark options={solid, colorDRSD, fill=white, mark size=1.5pt}]
   table[row sep=crcr]{%
3.89162 0.0150249\\
3.97079 0.0089769\\
4.04995 0.00249982\\
4.12912 0.000224489\\
4.20829 3.30875e-06\\
4.28745 9.83661e-09\\
 };
 \addlegendentry{DRSD$^{20}$}
 
  \addplot [color=colorDRSDplus20, line width=0.9pt, mark=diamond*, mark options={solid, colorDRSDplus20, mark size=1.5pt}]
   table[row sep=crcr]{%
3.89404 0.012168\\
3.97321 0.00502814\\
4.05238 0.000890619\\
4.13154 2.5053e-05\\
4.21071 2.1675e-07\\
4.28988 4.5108e-10\\
 };
 \addlegendentry{DRSD+$^{20}$}

\addplot [color=black, line width=0.9pt, dotted,  mark=square*, mark options={solid, black, fill=white, mark size=1pt}]
table[row sep=crcr]{%
3.89404 0.0101831\\
3.97321 0.00541643\\
4.05238 0.00137433\\
4.13154 7.59973e-05\\
4.21071 6.07155e-07\\
4.28988 1.39937e-09\\
};
\addlegendentry{ideal EaED$^{10}$}

\addplot [color=colorDRSDdeter, line width=0.9pt,   mark=square*, mark options={solid, colorDRSDdeter, mark size=1pt}]
table[row sep=crcr]{%
3.89404 0.0227486\\
3.97321 0.0200445\\
4.05238 0.016592\\
4.13154 0.0124535\\
4.21071 0.00553468\\
4.28988 0.00104541\\
4.36904 2.74613e-05\\
4.44821 2.1756e-07\\
4.52738 9.68796e-10\\
};
\addlegendentry{Determ. fill$^{10}$}

\addplot [color=colorDRSDdeter, line width=0.9pt, dashed,  mark=square*, mark options={solid, colorDRSDdeter, mark size=1pt}]
table[row sep=crcr]{%
3.89404 0.0231803\\
3.97321 0.0203785\\
4.05238 0.0161736\\
4.13154 0.0107693\\
4.21071 0.00364429\\
4.28988 0.000277544\\
4.36904 3.28072e-06\\
4.44821 5.21305e-09\\
};
\addlegendentry{Determ. fill$^{20}$}

  \addplot [color=colorDRSD, line width=0.9pt, mark=*, mark options={solid, colorDRSD, fill=white, mark size=1.5pt}]
   table[row sep=crcr]{%
3.03649 0.0351582\\
3.11983 0.0291384\\
3.20316 0.0213643\\
3.28649 0.0135962\\
3.36983 0.0069004\\
3.45316 0.00249271\\
3.53649 0.000454642\\
3.61983 5.12845e-05\\
3.70316 3.67004e-06\\
3.78649 2.16932e-07\\
3.86983 2.13473e-08\\
 };
 
   \addplot [color=colorDRSD, line width=0.9pt, dashed, mark=*, mark options={solid, colorDRSD, fill=white, mark size=1.5pt}]table[row sep=crcr]{%
3.03649 0.0296026\\
3.11983 0.022856\\
3.20316 0.0141033\\
3.28649 0.00503048\\
3.36983 0.00150155\\
3.45316 0.000165315\\
3.53649 2.1766e-05\\
3.61983 1.02171e-06\\
3.70316 6.83611e-08\\
 };

    \addplot [color=colorDRSDplus20, line width=0.9pt, mark=diamond*, mark options={solid, colorDRSDplus20, mark size=1.5pt}]table[row sep=crcr]{%
3.03649 0.0243675\\
3.11983 0.0160531\\
3.20316 0.00792504\\
3.28649 0.00318682\\
3.36983 0.000624824\\
3.45316 6.83286e-05\\
3.53649 3.63385e-06\\
3.61983 2.50748e-07\\
3.70316 1.33912e-08\\
 };

 \addplot [color=black, line width=0.9pt, dotted, mark=square*, mark options={solid, black, fill=white, mark size=1pt}]
   table[row sep=crcr]{%
3.03649 0.017426\\
3.11983 0.00927323\\
3.20316 0.00373593\\
3.28649 0.000849354\\
3.36983 0.000103219\\
3.45316 9.05872e-06\\
3.53649 3.61681e-07\\
3.61983 1.25232e-08\\
};

\addplot [color=colorDRSDdeter, line width=0.9pt,   mark=square*, mark options={solid, colorDRSDdeter, mark size=1pt}]
table[row sep=crcr]{%
3.03649 0.0435192\\
3.11983 0.0392959\\
3.20316 0.0343243\\
3.28649 0.0280065\\
3.36983 0.0188855\\
3.45316 0.00917174\\
3.53649 0.00378345\\
3.61983 0.000922528\\
3.70316 0.00011068\\
3.78649 9.59476e-06\\
3.86983 3.3231e-07\\
3.95316 2.2298e-08\\
};

\addplot [color=colorDRSDdeter, line width=0.9pt, dashed,  mark=square*, mark options={solid, colorDRSDdeter, mark size=1pt}]
table[row sep=crcr]{%
3.03649 0.043692\\
3.11983 0.0393567\\
3.20316 0.0339829\\
3.28649 0.0256018\\
3.36983 0.0165728\\
3.45316 0.00797991\\
3.53649 0.00222711\\
3.61983 0.000364812\\
3.70316 3.46303e-05\\
3.78649 1.48996e-06\\
3.86983 2.98821e-08\\
};

\node[ellipse,draw,dashed, anchor = center,at={(axis cs:3.4,1e-3)},minimum width = 2cm, 
    minimum height = 1cm] (e1)  {};
\node[font=\footnotesize] at (axis cs:3.6,1e-2) {$\CW_1$};

\node[ellipse,draw,dashed, anchor = center,at={(axis cs:4.2,1e-4)},minimum width = 1.7cm, 
    minimum height = 1cm] (e2)  {};
\node[font=\footnotesize] at (axis cs:4.4,1e-3) {$\CW_2$};
\end{axis}
\end{tikzpicture}  
	\caption{BER vs. $\Eb/\No$ (in dB) curve for product codes with component code $\CW_1$ and $\CW_2$ when decoded with various modifications of DRSD. The superscript of the labels denotes the maximum number of decoding iterations~$L$.}
	\label{fig:BERcurve255new}
	\vspace{-2ex}
\end{figure}
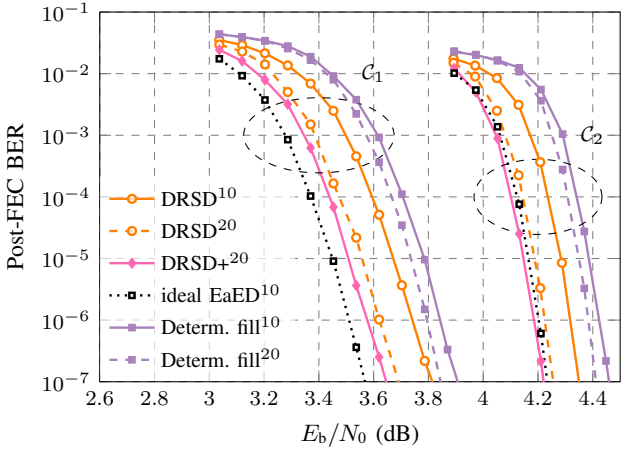

\subsection{The Importance of Random Filling Patterns}
In Fig.~\ref{fig:BERcurve255new}, additionally to the BER results after decoding with the proposed decoding scheme, we show the performance when using deterministic filling vectors  for the erased positions within the EaED. 
The purple curves depict the decoding performance using a DRSD but we replace the random filling pattern $\left(\pone,\ptwo\right)$ in Algorithm~\ref{alg:eaed} (line 5) with $\left(\vec{0},\vec{1}\right)$ as in~\cite[Sec. 3.8.1]{MoonBook}. This degrades the decoding performance by more than $0.1$\,dB for both PCs. Moreover, we observe that when using a deterministic value for $\pone$ and $\ptwo$, allowing $20$ iterations does not yield performance gains as large as we observe with DRSD. This is due to the fact that with identical filling patterns, row/column words which cause a decoding failure in early iterations will still cause a decoding failure if its EaEs are not corrected by the corresponding column/row code.

\subsection{Choice of Parameters}
In Sec.~\ref{sec:algo}, we gave the algorithm description with a predefined number of $32$ DRS representation levels and an initial DRS range of $[9,24]$. In this section, we provide some intuition and numerical simulation results which support this choice of parameters.

Figure~\ref{fig:pervslevel} shows the decoding performance of DRSD$^{20}$ with different DRS representation levels. We consider the number of levels in terms of the number of bits required for storage. All thresholds inside the algorithm are scaled relatively to the number of representation levels. If the number of representation levels is smaller than $32$, we observe a degradation of the decoding performance. However, if a smaller storage space is preferred, the DRSD algorithm can still be used with some loss in decoding performance. As using $64$ levels does not improve upon $32$ levels significantly, we choose $32$ levels for the proposed DRSD throughout this paper.

Next, we consider the range into which the initial DRS are distributed: first, the range should not be too small so that the channel output reliability information can be fully exploited. Second, it should not be too big either, otherwise,  the updating mechanism during  iterative decoding would have too little impact on the scores and thus cannot efficiently update the anchor bits. We exemplarily show the performance for some ranges $[a,b]$ in Fig.~\ref{fig:pervsinitialvalue} for 32 DRS representation levels. We observe that an initial range that is too big or too small degrades the decoding performance. Hence, we choose to use the initial range of $[9,24]$, as it is a good compromise between representing the soft channel output and leaving sufficient room for increasing and decreasing the scores.
\begin{figure}[t]
    \centering
    		\begin{tikzpicture}
		\pgfplotsset{grid style={dashed, gray}}
		\pgfplotsset{every tick label/.append style={font=\footnotesize}}
		
		\begin{axis}[%
		xshift=1.5cm,
		xmin=3.8,
		xmax=4.5,
		ymode=log,
		ymin=1e-7,
		ymax=0.1,
		yminorticks=true,
		axis background/.style={fill=white, mark size=1.5pt},
		xmajorgrids,
		ymajorgrids,
		yminorgrids,
		width=7.5cm,
        height=5.5cm,
		xtick={3.5,3.6,...,5.5},
		ytick={0.1,0.01,0.001,1e-4,1e-5,1e-6,1e-7,1e-8},
		xlabel={$\Eb/\No$ (dB)},
		ylabel={Post-FEC BER},
		label style={font=\small},
		legend cell align={left},
		legend style={anchor = south west,  at={(axis cs:3.8,1e-7)}, draw=none, fill opacity=1, text opacity = 1,legend columns=1,font=\footnotesize, row sep = 0pt}
		]
		\addplot [color=colorDRSD3bit, line width=0.9pt,dashed, mark=+, mark options={solid, colorDRSD3bit, mark size=1.5pt}]
		table[row sep=crcr]{%
			3.79404 0.0244156\\
			3.83363 0.0230626\\
			3.87321 0.0216098\\
			3.91279 0.0197203\\
			3.95238 0.0181563\\
			3.99196 0.0159761\\
			4.03154 0.0138187\\
			4.07113 0.0112919\\
			4.11071 0.00884985\\
			4.15029 0.00592251\\
			4.18988 0.00308066\\
			4.22946 0.00130945\\
			4.26904 0.000432872\\
			4.30863 9.7733e-05\\
			4.34821 1.60185e-05\\
			4.38779 1.94935e-06\\
			4.42738 1.62854e-07\\
			4.46696 7.11988e-09\\
		};
		\addlegendentry{8 levels (3 bits)}

		\addplot [color=colorDRSD4bit, line width=0.9pt, dashed, mark=diamond*, mark options={solid, colorDRSD4bit, mark size=1.5pt}]
table[row sep=crcr]{%
	3.79404 0.0221431\\
	3.83363 0.0201883\\
	3.87321 0.0183716\\
	3.91279 0.0161621\\
	3.95238 0.0136074\\
	3.99196 0.0107384\\
	4.03154 0.00796653\\
	4.07113 0.00471219\\
	4.11071 0.00222556\\
	4.15029 0.000894518\\
	4.18988 0.000175932\\
	4.22946 2.76015e-05\\
	4.26904 2.80974e-06\\
	4.30863 1.72873e-07\\
	4.34821 1.03135e-08\\
};
\addlegendentry{16 levels (4 bits)}
		\addplot [color=colorDRSD, line width=0.9pt, mark=*, dashed, mark options={solid, colorDRSD, fill=white, mark size=1.5pt}]
table[row sep=crcr]{%
	3.89162 0.0150249\\
	3.97079 0.0089769\\
	4.04995 0.00249982\\
	4.12912 0.000224489\\
	4.20829 3.30875e-06\\
	4.28745 9.83661e-09\\
};	\addlegendentry{32 levels (5 bits)}

		\addplot [color=colorDRSD6bit, line width=0.9pt, mark=*, mark options={solid, colorDRSD6bit, mark size=1.5pt}]
		table[row sep=crcr]{%
			3.79404 0.0205025\\
			3.83363 0.0186499\\
			3.87321 0.0158303\\
			3.91279 0.013578\\
			3.95238 0.0103172\\
			3.99196 0.00682852\\
			4.03154 0.00389876\\
			4.07113 0.0015503\\
			4.11071 0.000372068\\
			4.15029 6.13273e-05\\
			4.18988 6.59257e-06\\
			4.22946 4.81623e-07\\
			4.26904 2.69776e-08\\
		};	\addlegendentry{64 levels (6 bits)}

		\end{axis}
		\end{tikzpicture}
		
    \caption{DRSD$^{20}$ decoding performance for a  PC based on the $(255,238,2)$ even-weight BCH subcode with different  DRS representation levels.}
		\label{fig:pervslevel}  
\end{figure}
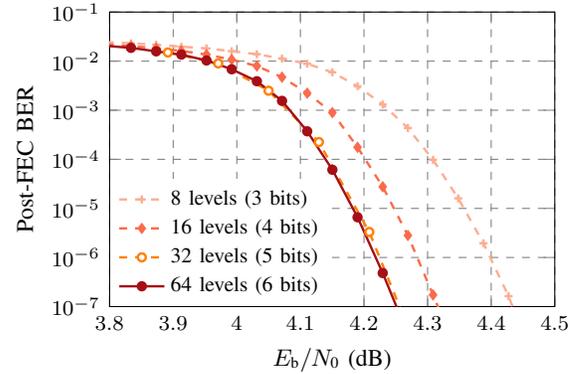
\begin{figure}[t]
    \centering
    		\begin{tikzpicture}
		\pgfplotsset{grid style={dashed, gray}}
		\pgfplotsset{every tick label/.append style={font=\footnotesize}}
		
		\begin{axis}[%
		xshift=1.5cm,
		xmin=3.8,
		xmax=4.5,
		ymode=log,
		ymin=1e-7,
		ymax=0.1,
		yminorticks=true,
		axis background/.style={fill=white, mark size=1.5pt},
		xmajorgrids,
		ymajorgrids,
		yminorgrids,
		width=7.5cm,
        height=5.5cm,
		xtick={3.5,3.6,...,5.5},
		ytick={0.1,0.01,0.001,1e-4,1e-5,1e-6,1e-7,1e-8},
		xlabel={$\Eb/\No$ (dB)},
		ylabel={Post-FEC BER},
		label style={font=\small},
		legend cell align={left},
		legend style={anchor = south west,  at={(axis cs:3.8,1e-7)}, draw=none, fill opacity=1, text opacity = 1,legend columns=1,font=\footnotesize, row sep = 0pt}
		]
		
		\addplot [color=colorDRSD, line width=0.9pt, dashed, mark=*, mark options={solid, colorDRSD, fill=white, mark size=1.5pt}]
		table[row sep=crcr]{%
			3.89162 0.0150249\\
			3.97079 0.0089769\\
			4.04995 0.00249982\\
			4.12912 0.000224489\\
			4.20829 3.30875e-06\\
			4.28745 9.83661e-09\\
		};
		\addlegendentry{[9,24]}

		\addplot [color=colorDRSD3bit, line width=0.9pt, mark=diamond*, dashed, mark options={solid, colorDRSD3bit, mark size=1.5pt}]
		table[row sep=crcr]{%
			3.79404 0.0235465\\
			3.83363 0.022187\\
			3.87321 0.0202168\\
			3.91279 0.0185946\\
			3.95238 0.0165881\\
			3.99196 0.0143266\\
			4.03154 0.011814\\
			4.07113 0.00906428\\
			4.11071 0.00629239\\
			4.15029 0.0036543\\
			4.18988 0.00159209\\
			4.22946 0.000520364\\
			4.26904 0.00011605\\
			4.30863 1.95314e-05\\
			4.34821 1.65169e-06\\
			4.38779 1.49323e-07\\
            4.42738 8.72942e-09\\
		};
		\addlegendentry{[12,19]}
		
		\addplot [color=colorDRSD6bit, line width=0.9pt, dashed,mark=*, mark options={solid,colorDRSD6bit, mark size=1.5pt}]
		table[row sep=crcr]{%
			3.79404 0.0226909\\
			3.83363 0.0210953\\
			3.87321 0.0195378\\
			3.91279 0.0178094\\
			3.95238 0.0160189\\
			3.99196 0.0141789\\
			4.03154 0.0126171\\
			4.07113 0.010438\\
			4.11071 0.00847876\\
			4.15029 0.00556342\\
			4.18988 0.00402004\\
			4.22946 0.00214697\\
			4.26904 0.000926004\\
			4.30863 0.000317746\\
			4.34821 7.47364e-05\\
			4.38779 1.43782e-05\\
			4.42738 2.19006e-06\\
			4.46696 2.71505e-07\\
			4.50654 2.92894e-08\\
		};
	\addlegendentry{[0,31]}
		
		\end{axis}
		\end{tikzpicture}
		
    \caption{DRSD$^{20}$ decoding performance for a PC based on the $(255,238,2)$ even-weight BCH subcode with different initial DRS ranges $[a,b]$.}
		\label{fig:pervsinitialvalue}  
\end{figure}
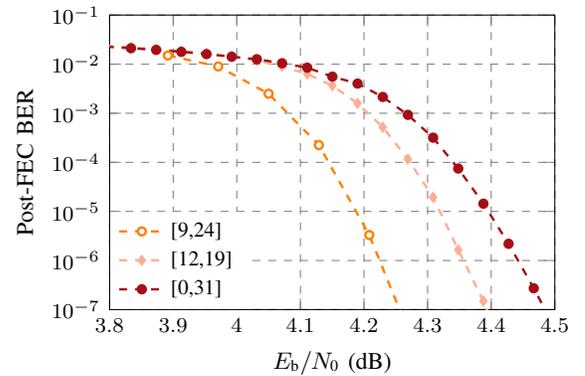
\begin{figure*}[tb]	
	\centering
	\includegraphics{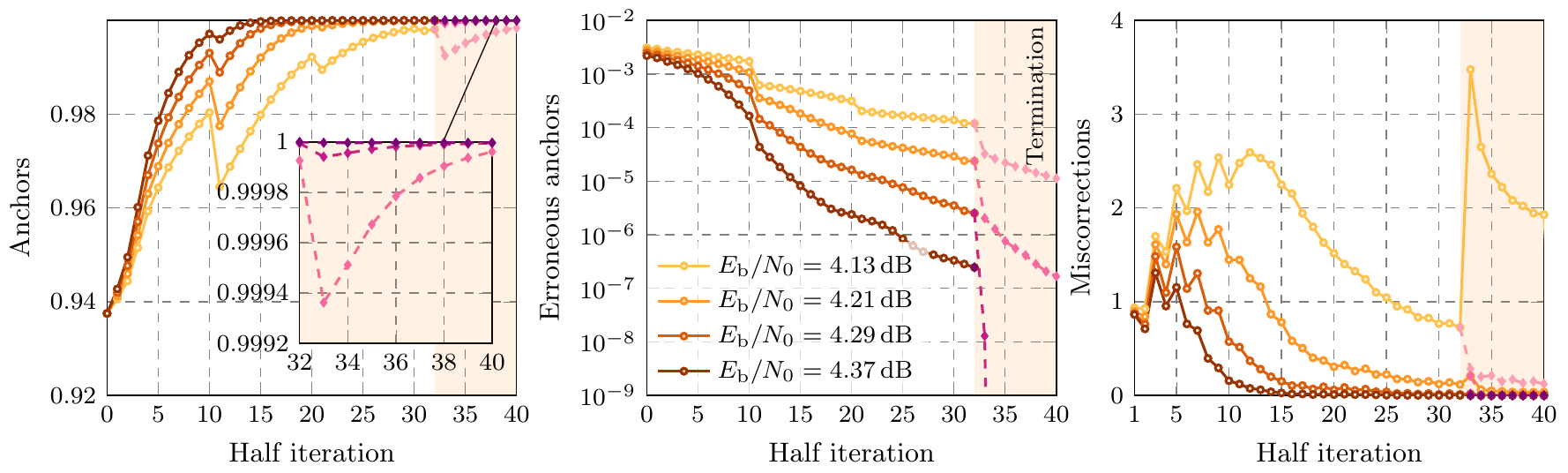}
	\vspace{-2ex}
	\caption{Percentage of marked anchor bits, percentage of wrongly marked anchor bits, and average number of miscorrections in every iteration in the decoding of a $(255,238,2)$ even-weight BCH subcode for various $\Eb/\No$. 
    The solid oranges lines stand for DRSD$^{20}$ and the dashed pink lines for DRSD+$^{20}$. }
	\label{fig:num_anchor}
    \vspace{-2ex}
\end{figure*}

\section{Heuristic Performance Analysis}
\label{sec:analy}

In this section, we investigate the iterative decoding process to show the connection between anchor bits and miscorrections. This also illustrates the advantage of the DRSD+ termination variant. Figure~\ref{fig:num_anchor} shows the fraction of anchor bits, erroneous anchor bits, and miscorrections during the decoding of a PC with component code $\CW_2$ and $L=20$ iterations. The quantities are provided for each half iteration which is defined as the processing of decoding all the rows \emph{or} all the columns of a PC block once. The values are calculated by averaging the respective quantities during the Monte-Carlo simulations which produced Fig.~\ref{fig:BERcurve}.

The leftmost subfigure depicts the fraction of anchors in each iteration and the middle subfigure depicts the fraction of wrongly marked anchor bits for different values of $\Eb/\No$ (the 0-th half iteration stands for the state after initialization and before iterative decoding). According to Sec.~\ref{sec:simu}, we set the initial anchor threshold to $\Ta=9$. Since there are 16 representation levels (in the interval $[9,24]$) for the initial DRSs, the initial fraction of anchor bits (DRS $>\Ta$) is always around $15/16\approx 93.8\%$. The number of erroneous anchor bits decreases as $\Eb/\No$ increases. The ratio of anchor bits increases during iterative decoding due to the raise of DRSs for the bits that are successfully decoded. At the same time, the DRS of the bits where the row and column decoders disagree is reduced. This reduces the number of erroneous anchor bits. $\Ta$ is increased by 1 after every 10 half-iterations, corresponding to the drop in both overall and erroneous anchor bits. As the last 4 iterations of DRSD are plain iEaED and the DRSs are not used, the solid lines stops after 32 half-iterations.

The rightmost subfigure in Fig.~\ref{fig:num_anchor} shows the number of miscorrections in each half iteration. For a poor $\Eb/\No$ with a high post-decoding BER, the DRS updates fail to raise the ratio of anchor bits and decrease the ratio of erroneous anchor bits sufficiently. Consequently, the number of miscorrections is high and even increases during decoding (due to error propagation). However, in the waterfall region, our decoder effectively prevents miscorrections. When $\Eb/\No\geq\;$4.29\,dB, nearly all decoded bits are marked correctly as anchor bits after a few iterations, resulting in almost no miscorrections, whereas a plain iEaED without miscorrection detection will have more than a hundred miscorrections in each iteration given the same $\Eb/\No$ value, as we observed during experiments.

As shown in Fig.~\ref{fig:num_anchor}, the number of miscorrections in the last few plain iEaED decoding iterations is relatively high. However, we observe that without these termination iterations, the decoding performance suffers from a high error floor due to erroneous anchor bits. The DRSD+ termination option removes almost all erroneous anchors with the high anchor threshold $\Ta^*$. This reduces the number of miscorrections in the last iterations while also eliminating the effect of erroneous anchor bits, leading to the improved decoding performance.

\section{Complexity Analysis}
\label{sec:complexity}
The proposed decoder does not require soft message passing. 
The \ac{DRS} register sends a binary message to the \ac{EaED} indicating whether a bit is an anchor or not. The \ac{EaED} signals to the \ac{DRS} register if DRSs should be increased or decreased. During EaE decoding, ternary messages are passed.

The major computational overhead of the proposed DRSD compared to \ac{iBDD} is the usage of \ac{EaED}, where the presence of erasures in a component word entails two \ac{BDD} steps. This increases the overall number of \ac{BDD} steps, especially in the first few iterations, before the erasures are resolved. Words without erasures can be decoded with simple \ac{BDD}.
The additional storage for \acp{DRS} is relatively small as the \acp{DRS} are stored using 5-bit integers.

\begin{figure}[tb]	
	\centering
	\includegraphics{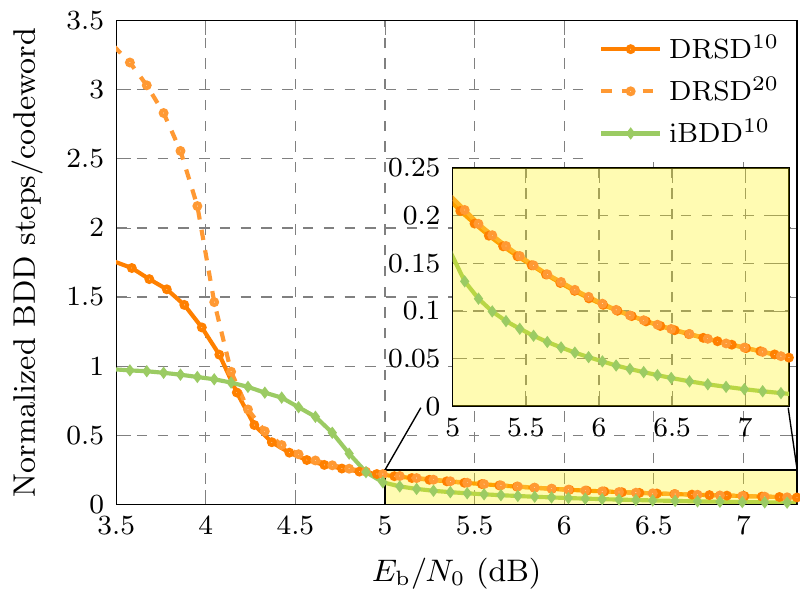}
	\caption{Normalized number of BDD steps in decoding of a $(255,238,2)$ even-weight BCH code-based PC with \ac{iBDD} and \ac{DRSD}.}
    \label{fig:num_decodings}
    \vspace{-2ex}
\end{figure}
As the main cause of the computational overhead is the increase in the number of BDD steps, we show in Fig.~\ref{fig:num_decodings} an exemplary comparison of the BDD steps in the decoding of a PC with component code $\CW_2$. See Fig.~\ref{fig:BERcurve} for the decoding performance of this code. We record the number of BDD steps used in three decoding schemes. One is the conventional iBDD$^{10}$, which serves as a baseline. Then we plot the results of DRSD$^{10}$ and DRSD$^{20}$ respectively. All the results are normalized by $(255\cdot 20)$, which is the total number of BDD steps in the iBDD$^{10}$ decoding process. Two facts cause the normalized value to be less than $1$: one is early termination due to the decoding success of the entire PC block. Another one is that we do not count the decoding of words with zero-syndrome, as the actual decoding algorithm does not need to be carried out. In the low $\Eb/\No$ region, the DRSD decoder requires a relatively large number of BDD steps. However, for larger $\Eb/\No$, the number of decodings steps 
is within the same order of magnitude as for iBDD. We also observe that the effective number of BDD steps in the DRSD$^{20}$ scheme increases only slightly compared to DRSD$^{10}$. Other soft-aided decoding schemes such as the SABM-SR decoder~\cite{liga2019novel} also require a number of BDD steps that is likely within the same order of magnitude than DRSD.

\section{Conclusion}
We proposed a novel hybrid decoding scheme for PCs and simulation results show that the proposed \ac{DRS} decoder significantly improves the decoding performance of the conventional hard decision \ac{iBDD} for \acp{PC} with near miscorrection-free EaE decoding. For two exemplary PCs of rate $0.87$ and rate $0.78$, DRSD closes roughly $80\%$ of the gap between hard-decision iBDD and soft-decision TPD which outperforms other soft-aided \ac{HDD} schemes. Complexity analysis shows that the decoding complexity is similarly low as for conventional \ac{HDD} iBDD decoder. The high \acp{NCG} make this scheme a promising candidate for future low-complexity optical communication systems. 

One of the open questions is the error floor of the proposed DRSD. With a proper choice of the parameters, we conjecture that the proposed DRSD should not have a worse error floor than a plain iterative EaE decoder, which has been investigated in~\cite{soma2021errors} for PCs with relatively small block lengths. The reason is that DRSD does not introduce additional miscorrections compared with the plain iterative EaE decoder. However, the EaE decoder used in this work is different from the EaE decoder studied in~\cite{soma2021errors}. Therefore, accurately calculating the error floor for decoding the PCs with large block lengths that are used in this work remains difficult and is left as future work.

\section*{Acknowledgment}
The authors would like to thank Alex Alvarado and Alireza Sheikh for providing the numerical BEE-PC simulation results used in Fig.~\ref{fig:BERcurve}.

\end{document}